\documentclass[12pt,a4paper]{article}
\usepackage{epsfig}
\newcommand{\be}{\begin{equation}}
\newcommand{\ee}{\end{equation}}
\newcommand{\ba}{\begin{eqnarray}}
\newcommand{\ea}{\end{eqnarray}}

\newcommand{\btab}{\begin{tabular}}
\newcommand{\etab}{\end{tabular}}

\newcommand{\bit}{\begin{itemize}}
\newcommand{\eit}{\end{itemize}}

\def\NNLO{\hbox{\tiny NNLO}}
\def\overlap{\hbox{\tiny overlap}}

\def\NP{\hbox{\tiny NP}}
\def\HT{\hbox{\tiny HT}}

\catcode`\@=11 
\def \lsim {\mathrel{\mathpalette\@versim<}}
\def \gsim {\mathrel{\mathpalette\@versim>}}
\def\gappeq{\mathrel{\rlap {\raise.5ex\hbox{$>$}}{\lower.5ex\hbox{$\sim$}}}}
\def\lappeq{\mathrel{\rlap{\raise.5ex\hbox{$<$}}{\lower.5ex\hbox{$\sim$}}}}
\def\@versim#1#2{\vcenter{\offinterlineskip
\ialign{$\m@th#1\hfil##\hfil$\crcr#2\crcr\sim\crcr } }}
\catcode`\@=12 

\newcommand{\mycomm}[1]{\hfill\break $\phantom{a}$\kern-3.5em{\tt===$>$ \bf
#1}\hfill\break}
\newcommand{\mycommA}[1]{\hfill\break $\phantom{a}$\kern-3.5em{\tt   $>$ \bf
#1}\hfill\break}
\renewcommand{\thefootnote}{\fnsymbol{footnote}}
\newcommand{\ysl}{\mbox{$y$\hspace{-0.5em}\raisebox{0.1ex}{$/$}}}

\newcommand{\psl}{\mbox{$p$\hspace{-0.4em}\raisebox{0.1ex}{$/$}}}

\def\MSbar {\hbox{$\overline{\hbox{\tiny MS}}\,$}}

\def \as {\relax\ifmmode\alpha_s\else{$\alpha_s${ }}\fi}

\usepackage{epsfig}
\title{essai}
\author{}
\date{\today}

\begin{document}

\begin{titlepage}
\begin{flushright}
{\small CERN-TH/2002-294}\\
{\small October, 2002}

\end{flushright}
\vspace{.13in}

\begin{center}
{\Large {\bf The interplay between Sudakov resummation,}}\\
{\Large {\bf renormalons and higher twist}} \\
{\Large {\bf in deep inelastic scattering}}

\vspace{.4in}

{\bf E.~Gardi}$\;^{(1,2)}$\, and\, {\bf R.G. Roberts}$\;^{(1,3)}$

\vspace{0.25in}

$^{(1)}$ TH Division, CERN, CH-1211 Geneva 23, Switzerland\\
\vspace*{10pt}
$^{(2)}$ Institut f\"ur Theoretische Physik, Universit\"at Regensburg,\\
D-93040 Regensburg, Germany\\
\vspace*{10pt}
$^{(3)}$ Rutherford Appleton Laboratory,\\
Chilton, Didcot, Oxon, OX11 0QX, UK
\vspace{.4in}

\end{center}
\noindent  {\bf Abstract:} We claim that factorization implies that 
the evolution kernel, defined by the logarithmic derivative of the 
$N$-th moment of the structure function $d\ln F_2^N/d\ln Q^2$, receives logarithmically enhanced contributions 
(Sudakov logs) from a single source, namely the constrained invariant
mass of the jet. Available results from fixed-order calculations facilitate
Sudakov resummation up to the next-to-next-to-leading logarithmic
accuracy. We use additional all-order information on the physical
kernel from the large-$\beta_0$ limit to model the behaviour of
further subleading logs and explore the uncertainty in extracting
$\alpha_s$ and in determining the magnitude of higher-twist
contributions from a comparison with data on high moments. 
\vspace{.25in}
\end{titlepage}

\def\thefootnote{\arabic{footnote}}
\setcounter{footnote} 0

\section{Introduction}

The standard collinear factorization in deep inelastic scattering (DIS) at twist two is based on separating hard physics associated with scales of order of the momentum transfer $Q^2$ into coefficient functions, and softer physics associated with the structure of the target into operator matrix elements which have the interpretation of parton distribution functions.

Considering the kinematic limit $x\longrightarrow 1$ (where Bjorken $x$ is defined by $x\equiv -q^2/2pq=Q^2/2pq$), an additional, well-separated scale emerges, $W^2=Q^2(1-x)/x$  which represents the invariant mass of the hadronic system. This is the characteristic scale for dynamics of the  jet in the final state. 
When $W^2$ is much smaller than $Q^2$, perturbative corrections (Sudakov double logs) related to soft and collinear radiation
which form the jet dominate the coefficient functions. These corrections become so large that the perturbative expansion breaks down.
Consequently, they must be resummed to all orders. 
In standard analysis of structure function data the difficult large-$x$ region
is usually avoided simply by putting a lower cut on $W^2$. For example, in the MRST analysis~\cite{Martin:2002dr} a cut is put such that $W^2>12.5$ GeV$^2$.

\begin{figure}[htb]
  \begin{center}  
\epsfig{height=5cm,angle=0,file=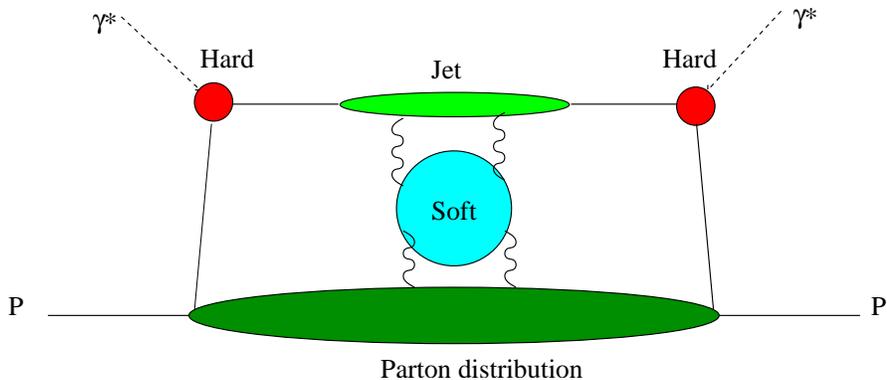}
\vspace*{-10pt}
  \end{center}
\caption{Factorization of DIS structure functions at large Bjorken $x$.}
\label{factorization}
\end{figure}

To describe the structure functions at large Bjorken $x$, it becomes natural to further factorize the process and treat the jet separately. Such factorization can indeed be accomplished~\cite{Sterman:1986aj}--\cite{Akhoury:1998gs}, 
facilitating the resummation of the logarithmic corrections. A schematic picture of large-$x$ factorization is shown in Fig.~\ref{factorization}. 

It is well known that higher-twist corrections are also enhanced at large $x$: they appear as powers of $\Lambda^2/W^2$ rather than $\Lambda^2/Q^2$. At sufficiently small $W^2$ the Operator Product Expansion (OPE) itself breaks down and needs to be resummed. Given the complexity of this expansion~\cite{JS}--\cite{Balitsky:1989bk},
the only hope this could ever be achieved is if a small subset of
matrix elements dominates at each twist, simplifying greatly the
expansion. There are two ways in which the {\em leading twist}
simplifies at large $x$. Firstly, the valence quarks dominate over the
gluons and the sea. Secondly, as explained above
(Fig.~\ref{factorization}) the dynamics of the outgoing jet,
associated with the scale $W^2$, decouples from the rest of the
process. The question then arises whether
any of this holds beyond the leading twist.

Recently, a first theoretical analysis of the large-$x$ limit of twist four has been performed~\cite{DIS}, suggesting that indeed, this kind of simplification occurs to all orders in the twist expansion, and thus its resummation at large $x$ may be possible.

Ref.~\cite{DIS} first establishes, in the concrete case of $F_2$, the cancellation of the infrared renormalon ambiguity at twist two with the quadratically divergent contribution of twist-four operators, associated with their mixing with twist two. Renormalons reflect the ambiguity in separating contributions of different twists. Independently of how this separation is implemented, within twist four {\em there is} a twist-two like ingredient, which is proportional to the twist two matrix elements. The question that arises is what is the significance of this twist-two like component compared with other ``genuine'' higher-twist contributions. Renormalon-based models for power corrections~\cite{Dasgupta:1996hh}--\cite{Meyer-Hermann:1996cy}
assume its dominance whereas other approaches may neglect it altogether.

The conjecture of~\cite{DIS} is that {\em at large $x$} the twist-two
like ingredient within the higher twist indeed dominates. Analysis of
the leading-order coefficient functions of {\em twist four} for $F_2$ 
and $F_L$ provides a physical picture which is consistent with this
conjecture. The final state which dominates the twist-four
contribution at large $x$ is the same as the twist-two final state: in
the case of $F_2$ this state contains a single energetic quark which initiates the jet.
Moreover, the analysis of~\cite{DIS} also reveals that the multi-parton initial states at twist four, which distinguish it from twist two, are dominated at large~$x$ by particular configurations which are twist-two like: a single quark in the initial state carries most of the momentum, whereas additional gluons carry small momentum fractions. Consequently, the corresponding twist-four matrix elements effectively depend only on the light-cone separation between the quark fields, similarly to twist two.

The conclusion is that factorization, similar to the one described by Fig.~\ref{factorization}, may hold beyond the perturbative level. The essential difference is that beyond twist two, gluons of virtuality of the order $W^2$ are exchanged between the jet and the remnants of the target. Their effect, to all orders in the OPE, can be taken into account through a single ``shape function''
of $N/Q^2$, which multiplies the twist-two contribution in moment space. This way, the conjecture of ref.~\cite{DIS} translates into a concrete understanding of how higher-twist effects should be parametrized at large $x$.

The need to resum power-corrections to all orders as a consequence of kinematic thresholds exists in a wide range of hard processes, including, for example, Drell-Yan pair production and event-shape distributions. In spite of the different nature of these processes, the analogy is quite useful. In particular, the appearance of a new non-perturbative distribution~\cite{KS}, the shape function, near the kinematic threshold, is common to all these processes, and so is the relation between the resummed perturbative distribution and the properties of this non-perturbative function~\cite{Gardi:2002yg,Thrust_distribution,DGE}. 

It should be stressed that the proposed non-perturbative factorization formula 
for DIS structure functions has not been derived from the OPE. It requires that the Sudakov resummed coefficient function of higher-twist contributions to $F_2$ would be the same as that of twist two. Moreover, the large-$N$ behaviour of higher-twist anomalous dimensions should coincide, asymptotically, with that of twist two. 
This highly non-trivial structure is yet to be verified by explicit~calculations. 

Having assumed the dominance of these higher-twist contributions which mix under renormalization with the leading twist (and are therefore proportional to the latter), renormalon {\em resummation} becomes absolutely essential for any higher-twist analysis. Indeed, the separation between the leading and higher twist is ambiguous. But, more importantly, prior to introducing any parametrization of power suppressed contributions one must be sure that any large perturbative corrections have already been taken into account~\cite{Gardi:1999dq,Thrust_distribution}. 
This concerns, in particular, running-coupling (or renormalon-related) perturbative corrections which are always parametrically larger than the corresponding ambiguity, and thus, by the previous assumption larger than the higher-twist contribution itself.
Failing to take these perturbative corrections into account, the parametrization of the higher twist becomes meaningless. For example, the values of the extracted ``higher twist'' parameters will strongly depend on the order of truncation of the perturbative expansion, the renormalization scale used, etc. 

The unresolved issue of higher twist in DIS structure functions dates far back.
It was understood very early on that higher-twist effects are especially important at large $x$. The limitations have always been both insufficient data and lack of theoretical understanding of this region.
In the recent years there has been much activity in this field, primarily owing to progress in perturbative calculations and resummation. Several groups have performed fits to data~\cite{Martin:1998np}--\cite{Alekhin:2002wp},
incorporating (or not) soft gluon resummation and parametrizing the higher twist in various ways. What lacks however, is a physical picture that stands behind these parametrizations. In particular, one is bound to ask eventually how the higher-twist parameters are related to operator matrix elements within the OPE, namely which multi-parton correlations are being measured. Ref.~\cite{DIS} suggests an answer\footnote{A previous suggestion with some similarity is that of ref.~\cite{Guo:1998ru}.}. More theoretical and experimental work is required to establish it. Here we make a further step in this direction. We concentrate mainly on perturbative aspects, constraining further the structure of the Sudakov exponent and examining where we stand on the long track to power accuracy.

As mentioned above, large-$x$ kinematics singles out very specific radiative corrections which dominate the coefficient function: these are the logarithmically-enhanced terms associated with the evolution of the jet. The jet function can be written as an exponential in moment space, where the exponentiation kernel at a given order can be systematically deduced from the standard loop expansion of the same order.  
To a first approximation (leading-log accuracy) the kernel simply corresponds to a single gluon emission calculation, whereas multiple emission is accounted for by the exponentiation (or solving the evolution equation with that kernel).
Refining the calculation of the kernel to next-to-leading logarithmic accuracy amounts to taking into account the leading contribution associated with the running of the coupling. A natural procedure to further improve the approximation to the kernel is a renormalon calculation with a single {\em dressed} gluon~\cite{Thrust_distribution,DGE}.
In this way some running-coupling effects are taken into account
to any logarithmic accuracy, making the exponent renormalization-scale invariant. Consequently the related power-corrections can be systematically parametrized, in accordance with~\cite{DIS}.

The essential difference between dressed gluon exponentiation (DGE), and other approaches to Sudakov resummation, is that it takes into account some all-order information on the exponentiation kernel itself. However, since the renormalon calculation is restricted to the large-$N_f$ (or, equivalently, large $\beta_0$) limit, it is necessary to combine it with what is known about the first few orders in the kernel from fixed-order calculations. One of the goals of this paper is to combine the large-$\beta_0$ result~\cite{DGE} for the Sudakov exponent of $F_2$ with the state-of-the-art knowledge
of the ${\overline {\rm MS}}$ anomalous dimension and coefficient function.
The latter practically\footnote{The non-Abelian contribution to the ${\overline {\rm MS}}$ anomalous dimension is still not available analytically as a function of $N$. However, there is a reliable estimate~\cite{vanNeerven:2000wp} of the leading $\ln N$ contribution based on the computation of specific moments~\cite{Larin:1993vu}.}
allows a complete~\cite{Moch:2002sn} next-to-next-to-leading logarithmic (NNLL) accuracy calculation, so together with the all-order large-$N_f$ result, it seems that the Sudakov exponent as a whole is well constrained. Nevertheless, we will see that without making further assumptions concerning the structure of the Sudakov exponent, there still is a significant uncertainty concerning the magnitude of subleading logs. This has important consequences for phenomenology.  

In this paper we apply, for the first time, DGE with the corresponding parametrization of higher-twist corrections to DIS data.
We restrict our attention here to the large-$x$ region, by analysing moments $N\geq 5$. The moment space analysis is advantageous from a theoretical point of view in several respects. First of all, it facilitates the implementation of target-mass corrections through the use of Nachtmann weights. In addition, it simplifies significantly the resummation formulae for the coefficient function as well as the higher-twist corrections.

The paper is organized as follows. The theoretical part (section 2) begins by recalling the non-perturbative factorization formula of~\cite{DIS}.
We then return to discuss factorization on the perturbative level and its consequences for resummation (section 2.1). We emphasize in particular
the fact that Sudakov logs in the physical evolution kernel are exclusively related to the jet function and not to the soft function. 
This translates into a more predictive formulation of Sudakov resummation than that commonly presented in the literature.
This discussion is followed by an all-order analysis of twist-two at large~$N$, first (section 2.2) for the physical kernel, and then (section 2.3) for the coefficient function. Sections 2.4 and 2.5 are devoted to combine 
the all-order large-$N_f$ information with the available coefficients in QCD and discuss in some detail the uncertainty involved in such a procedure. 
Section 3 is devoted to data analysis. Matching the DGE exponent into the known
~\cite{FRS}--\cite{vanNeerven:2001pe}
next-to-next-to-leading order (NNLO) result we obtain an improved perturbative prediction for the scaling violation at large $N$, which we use as a baseline for the study of power corrections.
The experimental moments $N=4$ -- $11$ of $F_2$ for the proton are calculated based on SLAC and BCDMS data. Each of the moments is fitted separately, to extract the valence quark distribution, $\alpha_s$ and the magnitude of the higher twist. We use the stability of the extracted value of $\alpha_s$ as a function of $N$ as a measure of the quality of our theoretical description of $F_2$, and compare the results to the standard NNLO analysis.  
In section 4 we summarize our conclusions.

\section{Factorization and the Sudakov exponent by DGE}  
  
\subsection{Factorization}

The {\em non-perturbative} factorization formula~\cite{DIS} for $F_2$ at large $N$ takes the form:
\begin{eqnarray}
\label{non_PT_fact}
F_2^N(Q^2) \equiv  \int_0^1 dx x^{N-2}{F_2}(x,Q^2)  
=H\left({Q^2}\right)\,J_N\left({Q^2};\mu_{F}^2\right)\,
q_N(\mu_{F}^2) \,J^{\NP}\left({N\Lambda^2}/{Q^2}\right).
\end{eqnarray}
We will assume that it holds up to corrections of order $1/N$.
Here, the first three factors correspond to the standard, perturbative  large-$x$ factorization~\cite{Sterman:1986aj}--\cite{Akhoury:1998gs}:
$H\left({Q^2}\right)$ is the hard part of the coefficient function depending on the momentum transfer $Q^2$,  $J_N\left({Q^2};\mu_{F}^2\right)$ is the Sudakov-resumed jet function depending primarily on the invariant mass of the jet $Q^2/N$, and $q_N(\mu_{F}^2)$ is the twist-two quark matrix element.
The last factor $J^{\NP}\left({N\Lambda}^2/{Q^2}\right)= 1+\kappa_1\,\frac{N\Lambda^2}{Q^2} +\cdots$, resums the dominant higher twist corrections to all orders~\cite{DIS}.

Some important clarifications are due concerning the perturbative\footnote{The reader in referred to~\cite{DIS} for discussion on power corrections.}  factorization formula.
First, the coefficients in the hard function $H\left({Q^2}\right)$ are finite at large $N$, so this function does not play any r\^ole in the large-$N$ limit.
The jet and the quark distribution are defined as follows:
\be
J\left(Q^2(1-\xi)/\xi;\mu_F^2\right)\,=\frac{1}{2\pi}\,{\rm Im}\,\int\, d^4\tilde{y} \,{\rm e}^{-i\,(q+zp)\tilde{y}}\,\left<0\left|{\rm Tr}\left(\Psi(\tilde{y})\overline{\Psi}(0)\psl\right)\right|0\right>,
\label{jet_def}
\ee
with $\xi\equiv x/z$, and 
\be
q\left(z;\mu_F^2\right)=\int_{-\infty}^{\infty}\,\frac{d(py)}{py}\, {\rm e}^{-i\,z\,py}\,\left<p\left|\overline{\Psi}(y)\ysl\Psi(0)\right|p\right> \quad\quad\quad\quad (y^2=0)
\label{q_def}
\ee
where the gauge in (\ref{jet_def}) and (\ref{q_def}) is chosen as $A_{-}=0$ and $A_{+}=0$, respectively. Here the lightcone `$+$' direction is defined by the incoming hadron $p$, and the lightcone `$-$' direction by the outgoing jet, $(q+xp)$. $J$ is simply a projection of the quark propagator in the axial gauge $pA=0$. At leading order $\Psi(\tilde{y})\overline{\Psi}(0)$ is the free quark propagator and $J=\delta(z-x)$; at higher orders
$J$ depends on $(q+zp)^2=Q^2(1-\xi)/\xi\simeq Q^2(1-\xi)$ but also, through the gauge fixing, on the hard scale $pq$. The latter dependence is unrelated
to logarithmically enhanced terms and it can be neglected. Corrections on this scale appear through the hard function~$H(Q^2)$.
 
\begin{figure}[htb]
  \begin{center}  
\epsfig{height=5cm,angle=0,file=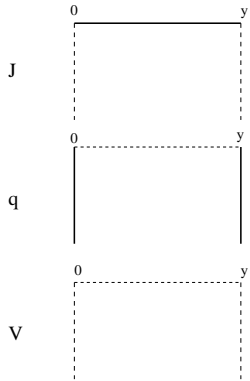}
\vspace*{-10pt}
  \end{center}
\caption{Factorization of $F_2$ at large $x$. Full and dashed lines stand for a dynamical quark and a Wilson line, respectively. The vertical lines correspond to the incoming lightcone direction $p$ and the horizontal line to the outgoing jet. The three pictures correspond to the jet, the quark distribution and the soft function, respectively.}
\label{JqV}
\end{figure}

In spite of the explicit dependence on the gauge, $J\left(Q^2(1-\xi)/\xi;\mu_F^2\right)$ and $q\left(z;\mu_F^2\right)$ are gauge invariant.
In case of different gauge choices, an appropriate path-ordered exponential should be introduced. For $J$ the path goes along the incoming quark direction~($p$), connecting the point~$\tilde{y}$ to infinity and then infinity to~$0$, whereas for $q$ the path goes along the jet direction between~$y$ and~$0$. The different paths are shown by the dashed lines in figure~\ref{JqV}. The functions in (\ref{non_PT_fact}) are defined in moment space, $J_N(Q^2;\mu_F^2)=\int_0^1dx\,x^{N-1}  \,J\left((1-x)Q^2;\mu_F^2\right)$ and $q_N(\mu_F^2)=\int_0^1dx\,x^{N-1}  \,q(x;\mu_F^2)$, 
where convolution in the momentum fraction simply becomes a product.

The $\mu_F^2$ dependence in (\ref{jet_def}) and (\ref{q_def}) is there to remind us that the definition of these matrix elements requires a prescription (factorization scheme and scale) to deal with infrared and ultraviolet divergence, respectively. Since $F_2$ is a physical quantity, this dependence must cancel out in the product. 

The standard formulation of large-$x$ factorization~\cite{Sterman:1986aj}--\cite{Akhoury:1998gs}
includes an additional soft function $V$:
\be
F_2(x,Q^2) =H\left({Q^2}\right)\, \, J\left((1-x)Q^2\right)\,\otimes
V\left(x;\mu_F^2\right)\,\otimes\,
q(x;\mu_{F}^2).
\label{standard_fact}
\ee
defined~\cite{Sterman:1986aj,CSS,Korchemsky:1988si,Korchemsky:1992xv} by the path-ordered exponential shown in the lower picture in figure~\ref{JqV}, which represents the DIS process in the Eikonal approximation.
The proof of factorization~\cite{Sterman:1986aj} relies on separating a generic Feynman diagram into subprocesses, each depending on a single scale. In inclusive DIS there are two external scales: $Q^2$ and $W^2=Q^2(1-x)/x\simeq Q^2/N$. The corresponding subprocesses are $H(Q^2)$ and $J(Q^2(1-x))$. Momentum scales which are parametrically smaller than $W^2$, e.g. ${\cal O}\left(Q^2(1-x)^2\right)$, are accounted for by the soft subprocess $V(x,\mu_F^2)$.

\begin{figure}[htb]
  \begin{center}  
\epsfig{height=5cm,angle=0,file=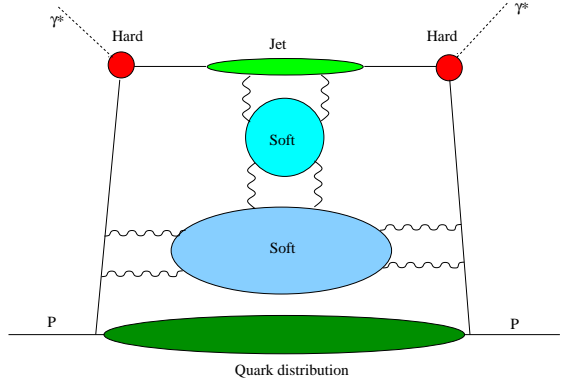}
\vspace*{-10pt}
  \end{center}
\caption{Factorization of $F_2$ at large $x$. The soft function is split between the jet and the quark distribution. Different separations amount to evolution of $q(x,\mu_F^2)$. In our formulation the upper `soft' blob is understood as part of the jet while the lower one as part of the quark distribution function.}
\label{factorization_js}
\end{figure}

It is important to note that $V(x,\mu_F^2)$ does not depend\footnote{Dependence of this function on $Q^2$ or on $Q^2/N^2$ is sometimes introduced through the factorization scale. This, however, can be misleading, and we will avoid it in this formal discussion.} on any external scale in the DIS process, and it is therefore possible and useful(!) to eliminate it from the factorization formula~\cite{DIS}.
Although not written explicitly in eq.~(\ref{standard_fact}), in order  
to define the soft function one actually performs two factorization procedures with two different scales~$\mu_{F_{1,2}}^2$: one in the definition of~$J\left((1-x)Q^2;\mu_{F_1}^2\right)$ and the other in the definition of~$q(x;\mu_{F_2}^2)$. This can be intuitively undersood from Fig.~\ref{factorization}, where the soft blob is attached to jet and to the quark distribution. 
In any case some component of the soft blob is included in the jet and some in the quark distribution. Since $V$ represents the evolution of the quark distribution as well as that of the jet function with $\mu_F^2$, it can be eliminated altogether by choosing the same scale in both. This is demonstrated in figure~\ref{factorization_js}.
The factorization formula is then 
\begin{eqnarray}
F_2^N(Q^2) &=&H\left({Q^2}\right)\,J_N\left({Q^2};\mu_{F_1}^2\right)\,
V_N\left(\mu_{F_1}^2,\mu_{F_2}^2\right)\,
q_N(\mu_{F_2}^2)\nonumber \\
&=&H\left({Q^2}\right)\,J_N\left({Q^2};\mu_{F_2}^2\right)\,
q_N(\mu_{F_2}^2).
\end{eqnarray}
where 
\[
J_N\left(Q^2;\mu_{F_2}^2\right)\equiv J_N\left(Q^2;\mu_{F_1}^2\right)\,
V_N\left(\mu_{F_1}^2,\mu_{F_2}^2\right).
\]

In the factorization formula~(\ref{non_PT_fact}) the soft function appears only implicitly through the dependence of $J_N(Q^2;\mu_F^2)$ and $q_N(\mu_F^2)$ on $\mu_F^2$. 
This function has some remarkable properties which will become relevant in the following. As discussed in ref.~\cite{Sterman:1986aj} (see section 7 there) and in~\cite{Korchemsky:1988si,Korchemsky:1992xv} its exponent contains just a single $\ln N$ at any order in the coupling. 
It thus follows from the factorization formula that apart from these $\mu_F^2$-related single logs, the only source of logarithmically enhanced contributions to the $F_2$ coefficient function is in the constraint on the invariant mass of the jet. This is true to all orders in perturbation theory.
As we discuss further in the next section, this implies, in particular, that the jet function $J\left((1-x)Q^2;\mu_F^2\right)$ alone determines the 
Sudakov exponent in the physical kernel $d\ln F_2^N(Q^2)/d \ln Q^2$.

We stress that the absence of additional sources of Sudakov logs is strictly 
related to the inclusive nature of structure functions. Note that, due to different kinematic circumstances, sensitivity to large-angle soft emission which is associated with the scale $Q^2/N^2$ does generate Sudakov double logs in other observables. Examples are provided by Drell-Yan pair production and event-shape distributions. 

Finally, in the notation of \cite{CT,Vogt:2000ci,Moch:2002sn}, our statement that the only source of Sudakov logs in the physical kernel is in the jet function means that $D_n^{\rm DIS}$ vanish to all orders. Indeed, recently, it was shown~\cite{Moch:2002sn}, based on explicit calculations in $\overline {\rm MS}$ that $D_2^{\rm DIS}=D_1^{\rm DIS}=0$, although the conviction that $D^{\rm DIS}$ vanishes to all orders was not made.

\subsection{The physical kernel}

It is instructive to formulate first Sudakov resummation in the physical kernel defined by the logarithmic derivative $d\ln F_2^N(Q^2)/d \ln Q^2$. 
We claim that the latter can be written, to any order in perturbation theory, as
\ba
\label{phys_kernel}
\frac{d\ln F_2^N(Q^2)}{d \ln Q^2}= \frac{d\ln J_N(Q^2;\mu_F^2)}{d \ln Q^2}=\frac{C_F}{\beta_0} \,\int_0^1\frac{x^{N-1}-1}{1-x}\,{\cal K}\left((1-x)Q^2\right)
\nonumber \\=
 \frac{C_F}{\beta_0} \,\int_0^\infty du \,T(u) \left(Q^2/\Lambda^2\right)^{-u}
\left(N^u-1\right)\,\Gamma(-u)\,B[{\cal K}](u),
\ea
where, as before, terms that are finite at large $N$ are neglected. Here we used the scheme invariant Borel representation~\cite{Grunberg}, 
\be
{\cal K}(\mu^2)= \int_0^{\infty} du\,B[{\cal K}](u)\,T(u)\left(\mu^2/\Lambda^2\right)^{-u},
\ee
and $T(u)$ is the Laplace transform of the coupling\footnote{Eq.~(\ref{phys_kernel}) is written in a renormalization scheme invariant way. Both $T(u)$ and the coefficients of $B[{\cal K}](u)$ would change depending on the renormalization group equation for the coupling, such that the product remains invariant.}, 
\be
A(Q^2)= \int_0^{\infty} du\,T(u)\left(Q^2/\Lambda^2\right)^{-u}.
\ee
In the following we will use the `t Hooft coupling $A\equiv\alpha_s(Q^2)\beta_0/\pi$, defined by
\be
\label{tHooft}
\frac{dA}{d\ln Q^2}=-A^2\,\left(1+\delta A\right);\,\,
\quad\quad\quad\delta\equiv \beta_1/{\beta_0}^2,
\ee
so
\be
\label{D}
T(u)=(u\delta)^{u\delta}\exp(-u\delta)/\Gamma (1+u\delta).
\ee
The function 
\[
B[{\cal K}](u)=1+\tilde{k}_1\frac{u}{1!}+\tilde{k}_2\frac{u^2}{2!}+\tilde{k}_3\frac{u^3}{3!}+\ldots
\] 
is the Borel representation of an $N$-{\em independent} effective charge, which is a function of the coupling having the expansion \[
{\cal K}=A+k_1 A^2+k_2A^3+\ldots\,.\] 
Our notation here is adopted for the large-$\beta_0$ limit, where $\delta=0$, $T(u)=1$ and $\tilde{k}_n=k_n$. Beyond this limit the general relation can be written as an expansion in~$\delta$,
\be
\label{k_tilde_relation}
\tilde{k}_n\,=\,k_n\,-\,\delta\,k_{n-1}\,n\,\left(\Psi(n+1)+\gamma_E-1\right)+{\cal O}(\delta^2).
\ee

Like the physical kernel itself, ${\cal K}$ is does not depend on any arbitrary  factorization procedure. Note that even if ${\cal K}$ is free of any infrared renormalons (as is the case in the large-$\beta_0$ limit, and perhaps also in the full theory -- see below), these do appear in ${d\ln F_2^N(Q^2)}/{d \ln Q^2}$ through the factor $\left(N^u-1\right)\Gamma(-u)$ which is associated with soft and collinear radiation under the large-$x$ phase-space constraint~\cite{DGE}. This factor is understood as follows:
gluon splitting off a quark gives rise to a $1/(1-x)_+$ singularity.
The phase-space constraint determining the maximal gluon virtuality is 
$k^2<W^2=Q^2(1-x)/x$. Using the Borel representation of the dressed gluon propagator, $k^2$ is raised to a power $-u$. Performing the phase-space integration, phase-space constraint promotes the $x\longrightarrow 1$ singularity to a cut: $(1-x)^{-1-u}$. Going to moment space and neglecting ${\cal O}(1/N)$ contributions one obtains: $(N^u-1)\Gamma(-u)$. 
As a result, at any positive integer value of $u$ where $B[{\cal K}](u)$ does not vanish, the physical kernel will have a renormalon singularity.

The all-order $B[{\cal K}](u)$ in the large-$N_f$ limit has been computed in~\cite{DGE}, and it is given by
\be
\left.B[{\cal K}](u)\right\vert_{\rm large\, \beta_0}=\frac12 \,\frac{\sin \pi u}{\pi u}\,{\rm e}^{Cu} \, \left[\frac1 {1-u}+\frac1{1-u/2}\right],
\label{large_beta_0_result}
\ee
where $C=5/3$ is the constant from the ${\overline {\rm MS}}$ renormalization of the fermion loop\footnote{We assume that $\Lambda$ in (\ref{phys_kernel}) is defined in this renormalization scheme, while the renormalization-group equation is~(\ref{tHooft}).}. 
Additionally, the first two coefficients $k_1$ and $k_2$ are known (see e.g.~\cite{Vogt:2000ci,vanNeerven:2001pe,Moch:2002sn}), by combining results for the coefficient function and the anomalous dimension in the ${\overline {\rm MS}}$ factorization scheme. The relations will be given below.   

Examining the (large $N_f$) renormalon structure of the physical kernel~(\ref{phys_kernel}) with (\ref{large_beta_0_result}) we deduce that non-perturbative corrections at large~$N$ appear as~\cite{DGE,DIS} a multiplicative factor in moment space, as summarized by eq.~(\ref{non_PT_fact}), with
\be
{J}^{\NP}(N \Lambda^2/Q^2)=\exp\left\{-\omega_1\frac{C_F}{\beta_0}\left(\frac{N\Lambda^2}{Q^2}\right)
-\frac12 \omega_2\frac{C_F}{\beta_0}\left(\frac{N\Lambda^2}{Q^2}\right)^2
\right\}.
\label{J_NP}
\ee
This power-correction model effectively resums the twist expansion, under the assumption that the higher-twist contributions which dominate at large $N$, are those which mix with the leading twist. 
Fixing the parameters $\omega_1$ and $\omega_2$ requires the knowledge of the 
dominant contribution to the matrix elements of the corresponding higher-twist operators~\cite{DIS}. 
Clearly, the non-perturbative contribution ${J}^{\NP}(N \Lambda^2/Q^2)$ has a meaning only within a given regularization prescription for the renormalons in the perturbative jet function~$J_N\left({Q^2};\mu_{F}^2\right)$.

In the large-$\beta_0$ limit ${\cal K}$ is free of any renormalon ambiguity. $B[{\cal K}](u)$ is an analytic function at any finite $u$, so renormalon singularities in the physical kernel are exclusively related to the end-point singularity at $x\longrightarrow 1$. If ${\cal K}$ is free of infrared renormalons also in the full theory, it follows\footnote{As usual~\cite{Mueller,Beneke}, the nature of infrared renormalon singularities is related with the anomalous dimension of higher-twist operators.} that at large~$N$ the anomalous dimension of higher-twist operators (which mix with twist two), coincides with that of the twist two, as was already anticipated\footnote{Let us recall that the anomalous dimension of twist three is known to coincide asymptotically with that of twist two~\cite{Braun:2001qx}.} in~\cite{DIS}. 
Another possibility not yet excluded is that beyond the large-$\beta_0$ limit,
$B[{\cal K}](u)$ would have infrared renormalons. For example, the poles at $u=1$ and $u=2$, which are exactly compensated by the sine factor in~(\ref{large_beta_0_result}), might become branch points. Such a cut structure would imply that the large $N$ asymptotics of higher-twist anomalous dimensions differ from that of the leading twist, and a more complicated pattern of power corrections at large~$N$ would emerge.

Finally, let us return to the last point we made on section 2.1. 
The absence an additional source of Sudakov logs of the form 
${\cal D}^{\rm DIS}((1-x)^2Q^2)$ in inclusive DIS (such a term was included, for example, in~\cite{Vogt:2000ci}) can also be understood from general consideration on power corrections. 
In  DIS the external Lorentz invariants are~$Q^2$ and~$W^2\sim Q^2/N$. On these grounds alone one expects power corrections to appear on either of these two scales and not on any other scale. For example, corrections such as $\Lambda^2 N^2/Q^2$ should not appear. On the other hand, had there been 
a function ${\cal D}^{\rm DIS}((1-x)^2Q^2)$ in eq.~(\ref{phys_kernel}), renormalons associated with the scale $Q^2/N^2$ would have appeared\footnote{The corresponding Borel integrand includes the factor~$(N^{2u}-1)\Gamma(-2u)$, so unless $B[{\cal D}](u)$ vanishes at all integer and half integer values of $u$, there would be power corrections on the scale $Q^2/N^2$.} upon performing the integration over~$x$, similarly to the situation in Drell-Yan production~\cite{Beneke:1995pq,DGE}. As we argued in section 2.1, the perturbative factorization of structure functions does not require to introduce such a function and this problem is avoided altogether.

\subsection{The jet function}

Going from the physical kernel to $\ln F_2^N$ itself, the differential equation~(\ref{phys_kernel}) should be supplemented by an initial condition. Within the OPE, the initial condition is defined as the matrix element of the twist-two quark operator with $N$ covariant derivatives, $q_N(\mu_F^2)$. The operator has a non-trivial renormalization, which takes the form
\be
\label{q_N_ev}
\frac{d\ln q_N(\mu_F^2)}{d \ln \mu_F^2}= -\,\frac{C_F}{\beta_0}\, \ln N\, {\cal A}
\ee
up to corrections which are finite at large~$N$.
The effective charge 
\[
{\cal A}=A+a_2A^2+a_3A^3+\ldots
\] 
is associated with the soft function $V$ in section 2.1. It corresponds to the $1/(1-x)_+$ singular part in the Altarelli-Parisi splitting function, which can be computed using the Eikonal approximation, or the Wilson lines shown in the lower picture in Fig.~\ref{JqV}. It has been noted that ${\cal A}$ is a rather universal object~\cite{Korchemsky:1988si,Korchemsky:1992xv}: it is the anomalous dimension of a Wilson line with a cusp. Upon integrating~(\ref{phys_kernel}) and (\ref{q_N_ev}) and using the factorization formula~(\ref{non_PT_fact})
to write $\ln J_N$
 as the difference between~$\ln F_2^N$ and~$\ln q_N$ (both regularized by the Borel variable~$u\neq 0$), we obtain the following jet function:
\ba
\label{J_N}
\ln J_N\left({Q^2};\mu_{F}^2\right)&=&
-\,\frac{ C_F}{\beta_0}\,\int_0^{\infty}\,du \,T(u)\,\left(Q^2/{\Lambda}^{2}\right)^{-u}
\nonumber \\ &&
\hspace*{10pt}\left[\Gamma(-u)\,\left(N^u-1\right) \,\frac{B[{\cal K}](u)}{u}\,
+\,\left(\mu_F^2/{Q}^{2}\right)^{-u} \frac{B[{\cal A}](u)}{u}\,\ln N\,\right], 
\ea
where, as before we use the scheme invariant Borel representation, where
\be
B[{\cal A}](u)=1+\tilde{a}_2\frac{u}{1!}+\tilde{a}_3\frac{u^2}{2!}+\ldots.
\ee
Note that the second term in the squared brackets cancels the $u=0$ singularity of the first, so in $\ln J_N\left({Q^2};\mu_{F}^2\right)$ the integral over $u$ exists. This is in contrast with $\ln q_N$ and $\ln F_2^N$, which are not perturbative entities. The dependence of $\ln J_N\left({Q^2};\mu_{F}^2\right)$ on the factorization scale $\mu_F$ and on the factorization scheme is guaranteed to cancel that of $q_N(\mu_{F}^2)$.

Eq.~(\ref{J_N}) summarises the perturbative jet function of $F_2$, namely the $\ln N$ enhanced contribution to the corresponding coefficient function.
It is written in an all-order, scheme-invariant way. 
There are various ways to define the Borel integral in (\ref{J_N}). In general, different regularizations of the renormalons differ by power terms of the form~(\ref{J_NP}). Therefore, once power terms are included it does not matter which particular regularization is chosen~\cite{Gardi:1999dq}: difference between different regularizations amounts to redefining~$\omega_i$. 

Here we choose to define (\ref{J_N}) by truncating the corresponding perturbative series 
\be
\ln J_N\left({Q^2};\mu_{F}^2\right)=
\,\frac{ C_F}{\beta_0}\sum_{k=0}^{\infty}\,\sum_{l=1}^{k+1} C_{k,l} \,(\ln N)^l\,   \,A^k,
\label{double_ser}
\ee
with $A\equiv \beta_0\alpha_s/\pi$, at the minimal term, i.e. at the point where the terms start increasing in magnitude. 
Even this can be done is several ways, which differ by power corrections.  One possibility would be to perform the sum over $l$ for a given $k$ and truncate the sum over $k$ at the minimal term. We will use another: let us rewrite the sum as in resummation with a fixed logarithmic accuracy,
\be
\ln J_N\left({Q^2};\mu_{F}^2\right)=
\,\frac{ C_F}{\beta_0}\sum_{m=0}^{\infty} g_m(\lambda) A^{m-1},
\label{lnJ_fla}
\ee  
where $g_m(\lambda)$ are functions of $\lambda=A\ln N$, and truncate the sum over $m$ at the minimal term.

It is straightforward to expand the Borel integrand in (\ref{J_N}), and derive explicit expressions for $g_m(\lambda)$.
Let us choose $\mu_F^2=Q^2$ and write $g_m(\lambda)$ as a series in $\delta=\beta_1/\beta_0^2$:
\be
g_m(\lambda)\simeq g_m^{(1)}(\lambda)+\delta\,g_m^{(2)}(\lambda)+\delta^2\,g_m^{(3)}(\lambda)+\cdots,
\label{delta_exp}
\ee
where first term appears in the large-$\beta_0$ limit (setting $T(u)=1$) and second and on are associated with integrating over the two-loop coupling using (\ref{D}). The expansion in powers of $\delta$ is useful since we will eventually discard higher powers of $\delta$, 
which enter for the first time at rather high orders (note that $\delta$ itself is not small $\delta(N_f=4)=0.7392$).
 
The functions $g_m^{(1)}(\lambda)$ and $g_m^{(2)}(\lambda)$ are given by 
\ba
\label{g_m}
g_m^{(1)}(\lambda)&=&\,r_m(\lambda)\,\tilde{p}_m/m!\,+\,\theta(m-1)\,\lambda\,
\,(\tilde{p}_m-\tilde{a}_{m+1})/m\\ \nonumber
g_m^{(2)}(\lambda)&=&\,\theta(m-1)\,\tilde{r}_{m}(\lambda)\tilde{p}_{m-1}/{(m-1)}!\,+\,\theta(m-2)\,\lambda\,(\psi(m)+\gamma_E-1) (\tilde{p}_{m-1}-\tilde{a}_{m}),
\ea 
where the coefficients $\tilde{p}_n$ (${\cal P}=A+p_1A^2+p_2A^3+\ldots$) are defined through the expansion of 
\be
B[{\cal P}](u)\equiv\Gamma(1-u)\,B[{\cal K}](u)=1+\tilde{p}_1\frac{u}{1!}
+\tilde{p}_2\frac{u^2}{2!}+\cdots.  
\label{P_def}
\ee
In (\ref{g_m}), aside from a linear term associated with the factorization procedure, the $\lambda$-dependence appears only through the functions
\ba
r_m(\lambda)&=&\sum_{n=0}^{\infty}\frac{(m+n)!}{(n+2)!}\,\lambda^{n+2}=\Gamma(m-1)\,\left[(1-\lambda)^{1-m}-\lambda (m-1)-1\right] \nonumber\\
\tilde{r}_m(\lambda)&=&\left(\frac{d}{dm} +\gamma_E-1\right)\,r_{m}(\lambda),
\ea
which are defined\footnote{For $m=0,1$ the limit should be taken carefully, yielding logarithms of $(1-\lambda)$.} for $m\geq 0$.
Note that both $r_m(\lambda)$ and $\tilde{r}_m(\lambda)$ are free of a linear term: their expansions start at order $\lambda^2$. 
These functions
have two generic properties \cite{Thrust_distribution,DGE,vanNeerven:2001pe} associated with their origin in an integral over the running coupling: they increase factorially with $m$ due to infrared renormalons and posses an increasing singularity at $\lambda=1$, corresponding to $Q^2/N\simeq\Lambda^2$, where the perturbative treatment looses its validity. 

Terms with $\delta^2$ enter first at NNLL accuracy ($m=2$):
\be
g_2^{(3)}(\lambda)=\frac12\,\frac{(\lambda+\ln(1-\lambda))^2}{1-\lambda}. 
\ee
The lowest order contribution here is $\Delta\ln J_N\left({Q^2};\mu_{F}^2\right)= (C_F/\beta_0)\,\frac18
\,\delta^2\,A^5\ln^4 N$, where we kept only the highest power of $\ln N$. Contrary to other terms of the same logarithmic accuracy (NNLL) this term has no large numerical coefficients. Normalizing by the magnitude of the leading term in $\ln J_N\left({Q^2};\mu_{F}^2\right)=(C_F/\beta_0)\,\frac12
\,A\ln^2 N$, and substituting $A\ln N=1$ (for a worst-case estimate) the corrections is of relative size of $(\delta A/2)^2$, i.e. $\sim 0.3$\% at $Q^2=10$ GeV$^2$. In the following we will work with exact NNLL resummation, supplemented by estimates of subleading logs (N$^3$LL and beyond, up to the minimal term in the series). As we will see below, estimates of subleading logs are anyway of limited accuracy. So there we will take into account just the first two terms in (\ref{delta_exp}), neglecting contributions ${\cal O}(\delta^2)$, which are tiny.
Note that dealing with N$^3$LL and beyond we will also be neglecting running-coupling effects associated with four-loop contributions to the $\beta$ function, as these corrections are neglected here in the relation between the `t Hooft coupling and ${\overline {\rm MS}}$.

It is useful to relate the Borel representation $B[{\cal K}](u)$ to the commonly used functions ${\cal A}$ and ${\cal B}$. The relation is:
\be
B[{\cal K}](u)=B[{\cal A}](u)\,-\,u\,B[{\cal B}](u)
\label{KAB_relation}
\ee
or
\be
{\cal K}={\cal A}\,+\,\frac{d{\cal B}}{d\ln Q^2}.
\ee
If we write ${\cal B}=b_1 A+b_2 A^2+b_3 A^3+\ldots$ with
\be
B[{\cal B}](u)=\tilde{b}_1+\tilde{b}_2\frac{u}{1!}+\tilde{b}_3\frac{u^2}{2!}+\ldots.
\ee 
then $\tilde{k}_n=\tilde{a}_{n+1}-n\tilde{b}_n$. Finally, it is straightforward to write the relation with $a_n^{\MSbar}$ and $b_n^{\MSbar}$ defined in the $\overline {\rm MS}$ renormalization scheme\footnote{Note that our notation for the coefficients of ${\cal A}$ and ${\cal B}$
is somewhat different from~\cite{vanNeerven:2000wp,Moch:2002sn}. The relation is $A_n=C_F 4^n\beta_0^{n-1}a_n^{\MSbar}$ and similarly for $B_n$.}.
For the first two orders (which are known in QCD) we have
\ba
\begin{array}{lll}
 \tilde{a}_1=a_1=1\hspace*{150pt}& \tilde{b}_1=b_1=-3/4\quad\quad&\tilde{k}_0=1\\
 \tilde{a}_2=a_2=a_2^{\MSbar}& \tilde{b}_2=b_2=b_2^{\MSbar}&\tilde{k}_1=\tilde{a}_2-\tilde{b}_1\\
 \tilde{a}_3=a_3-\delta a_2= a_3^{\MSbar}+\delta_2^{\MSbar}-\delta a_2^{\MSbar}&                         &       \tilde{k}_2=\tilde{a}_3-2\tilde{b}_2,
\end{array}
\label{kab_tilde_rel}
\ea
where $\delta_2^{\MSbar}=\beta_2^{\MSbar}/\beta_0^3$.
The $\overline {\rm MS}$ coefficients are given by~\cite{CT,CMW,Korchemsky:1992xv,Moch:2002sn}, 
\ba
\label{MSbar_coef}
a_2^{\MSbar}(N_c,\beta_0)&=&\frac53+\left(\frac13-\frac{\pi^2}{12}\right)\frac{C_A}{\beta_0},\\
 a_3^{\MSbar}(N_c=3,\beta_0)&=& -\frac13+\frac{1}{\beta_0}\left(\left(\frac{55}{16}-3\zeta_3\right)C_F+\left(\frac{253}{72}-\frac{5}{18}\pi^2+\frac72\zeta_3\right)C_A\right)-\frac{19.5\pm 0.2}{\beta_0^2}\nonumber \\ \nonumber
b_2^{\MSbar}(N_c,\beta_0)&=& -\frac{247}{72}+\frac16 \pi^2+\frac{1}{\beta_0}\left(\left(-\frac3{32}-\frac32\zeta_3+\frac18\pi^2\right)C_F
+\left(\frac{-73}{144}+\frac{5}{2}\zeta_3\right)C_A\right),
\ea
where the purely non-Abelian ingredient in $a_3^{\MSbar}$ is still not known exactly. However, a reliable estimate~\cite{vanNeerven:2000wp} was extracted from the first few moments~~\cite{Larin:1993vu} for $N_c=3$, allowing us to write the above.

\subsection{Naive non-Abelianization}

As explained in the introduction, our task is to take into account all the perturbative information available on $\ln J$, namely the large-$N_f$ limit and the first few coefficients known in QCD. 
The first question is, of course, to what extent does ``naive non-Abelianization'' provide an estimate of the coefficients.  
The answer, as reflected in the numerical values of the first few coefficients for $N_c=3$,
\ba
\label{MSbar_coef_Nc3}
a_2^{\MSbar}&=&1.667-\frac{1.467}{\beta_0},\\
 a_3^{\MSbar}&=& -0.333+\frac{14.714}{\beta_0}-\frac{19.5\pm 0.2}{\beta_0^2}\nonumber \\ \nonumber
b_2^{\MSbar}&=& -1.786+\frac{6.6104}{\beta_0},
\ea
is quite clear: the terms that are leading in the large-$\beta_0$ limit, are {\em not at all} dominant for the realistic value $\beta_0(N_f=4)\simeq 2$. 
\begin{table}[htb]
\begin{center}
\begin{tabular}{|l|l|l|l|l|}
\hline
$n$ &$a_{n+1}^{\MSbar}$(large $\beta_0$)& $b_n^{\MSbar}$ (large $\beta_0$)& $k_n^{\MSbar}$ (large $\beta_0$)& $p_n^{\MSbar}$ (large $\beta_0$)\\
\hline
0 &   +1    &         0        &      +1       & 1        \\ 
1 &  +1.667&         $-$0.75   &     +2.417   & 2.994    \\
2 &  $-$0.333&     $-$1.786    &   +3.238     & 8.006    \\
3 &  $-$2.071&      +0.425     &  $-$3.347    & 22.046   \\
4 &  $-$1.093&      +7.647     &  $-$31.683   & 75.214   \\
5 &  +0.325&        +15.591    &  $-$77.629   & 343.566   \\
6 &  +0.477&       +1.98       &  $-$11.404   & 1983.269  \\
7 &  +0.107&       $-$79.714   &  +558.106    & 13618.4   \\
8 &  $-$0.0487&    $-$223.285  &  +1786.23    & 107843.3  \\
9 &   $-$0.0296&    $-$83.078  &  +747.675    & 965598.1  \\
10&   $-$0.00236&    +1319.7   &   $-$13197.2 & 9631151.8 \\
11&  +0.00224 &      +4245.12  &   $-$46696.3 & 105806143.15     \\
12&  +0.000694   &    +1756.42 &  $-$21076.0  & 1268854143.8     \\
\hline                                        
\end{tabular}\\
\vspace*{30pt}
\begin{tabular}{|l|l|l|l|l|}
\hline
$n$ &\,\,$a_{n+1}^{\MSbar}$\quad\quad\quad\quad\quad&\,\, $b_n^{\MSbar}$ \quad\quad\quad\quad\quad&\,\, $k_n^{\MSbar}$\quad\quad\quad\quad\quad & \,\,$p_n^{\MSbar}$\quad\quad\quad\quad\quad\\
\hline
0 &   +1    &         0        &      +1         & 1             \\ 
1 &  +0.962&         $-$0.75   &     $+$1.712       & 2.289      \\
2 &  +2.232 (2.934)&   $+$1.387  &   $+$0.0115 (0.714) & 4.393 (5.095) \\
\hline                                        
\end{tabular}
\caption{The coefficients of ${\cal A}$, ${\cal B}$, ${\cal K}$ and ${\cal P}$ defined in (\ref{q_N_ev}), (\ref{KAB_relation}), (\ref{phys_kernel}) and (\ref{P_def}), respectively in the large-$\beta_0$ limit (upper table) and in full QCD with $N_f=4$ (lower table). In the latter, for $n=2$ the coefficient in the `t Hooft scheme is given in parenthesis.\label{cf_large_beta0}}  
\end{center}
\end{table}
This is also reflected in Table~\ref{cf_large_beta0} (compare, for example, the $n=2$ values in the lower and upper tables). 
One simply cannot use the large~$\beta_0$ 
limit directly to provide an estimate\footnote{This negative statement
does not apply to physical quantities such as the first few moments
of the structure functions. In this case estimates based on ``naive
non-Abelianization'' are reasonable~\cite{Maxwell,Beneke,Kataev}.}
of ${\cal A}$, ${\cal B}$ or ${\cal K}$
because these functions do not contain renormalons, and thus, there is no reason that the large-$\beta_0$ terms will have large coefficients. In fact, it is known that the series for the cusp anomalous dimension ${\cal A}$ {\em converges} in this limit into~\cite{Gracey,Beneke:1995pq},
\be
\left.{\cal A}\right\vert_{{\rm large}\, \beta_0} = \frac{\sin \pi A}{\pi}\,\,\frac{\Gamma(4+2A)}{6\Gamma(2+A)^2}\,+\,\cdots,
\label{G_large_Nf}
\ee
where the ${\overline {\rm MS}}$ scheme is used, and the dots stand for terms that are subleading at large~$\beta_0$.
It follows from~(\ref{large_beta_0_result}) and~(\ref{KAB_relation}) that ${\cal B}$, is renormalons-free as well\footnote{Although the coefficients of ${\cal B}$ are known to all orders, a closed-form expression as a function of the coupling is not known.}. Thus, also here ``naive non-Abelianization'' is not expected to apply.   

The status of~${\cal P}$ is somewhat different. It does contains renormalons so here the terms that are leading in the large-$\beta_0$ limit are expected to dominate at large orders. This does not necessarily imply that ``naive non-Abelianization'' estimates will be reliable at low orders, but it is expected that the general trend of increase characterizing $p_n$ in the large-$\beta_0$ limit right from the start, will persist in the full theory. The first few entries in the lower Table~\ref{cf_large_beta0} are consistent with this expectation.

While the large-$\beta_0$ coefficients themselves cannot be used to estimate
the actual QCD ones, more general properties may be deduced. As reflected in Table~\ref{cf_large_beta0} neither~${\cal A}$ nor~${\cal B}$ has a regular sign pattern. The sign is flipped every three or four terms owing to the $\sin \pi u$ function in (\ref{large_beta_0_result}) associated with taking a final-state cut. Another property is that in the large-$\beta_0$ limit the coefficients of ${\cal A}$ decrease with order while those of ${\cal B}$ have a general tendency to increase. Starting at $n\gsim 4$ there is a clear hierarchy $\left|a_{n+1}\right|\ll \left|k_n\right|$, i.e. the physical kernel is dominated by ${\cal B}$: $k_n=a_{n+1}-nb_n\simeq nb_n$. An even stronger hierarchy exists between ${\cal A}$ and ${\cal P}$: $\left|a_{n+1}\right|\ll \left|p_n\right|$, as the latter contains renormalons.

Finally, to make concrete predictions for the subleading logs in~(\ref{g_m}), the knowledge of~$\tilde{p}_n-\tilde{a}_{n+1}$ and $\tilde{p}_n$ at higher  orders ($n\geq 3$) is required. However, from the discussion above it is expected that at higher orders $\left|\tilde{a}_{n+1}\right|\ll \left|\tilde{p}_n\right|$, and thus the knowledge of ${\cal A}$ to higher orders is absolutely inessential. As usual, statements about large orders do not necessarily apply for any $n\geq 3$. 
\begin{table}
\begin{center}
\begin{tabular}{|l|l|l|}
\hline
$n$ &\,\,$\tilde{a}_{n+1}$\quad\quad\quad\quad\quad& \,\,$\tilde{p}_n$\quad\quad\quad\quad\quad\\
\hline
0 &   1    &          1      \\ 
1 &  0.962 &           2.289         \\
2 &  2.223 &          3.403         \\
\hline                                        
\end{tabular}
\caption{The coefficients of $B[{\cal A}](u)$ and $B[{\cal P}](u)$ in QCD with $N_f=4$.\label{ap}}  
\end{center}
\end{table}
Indeed, as shown in Table~\ref{ap} the expected hierarchy is not realised for $n<3$. Yet, the $n=3$ entries in
Table~\ref{cf_large_beta0} are reassuring: they suggest an order of magnitude difference between $\tilde{p}_3$ and $\tilde{a}_4$.
We will therefore orient ourselves in the next section to estimating ${\cal K}$ and thus ${\cal P}$ in full QCD while for ${\cal A}$ we will simply use the first known coefficients ($\tilde{a}_2$ and $\tilde{a}_3$) supplemented by the large-$\beta_0$ contribution (although the latter does not dominate!) for the higher orders. An alternative is to neglect $\tilde{a}_n$ for $n\geq 4$ altogether.

\subsection{Going beyond the large-$\beta_0$ limit}    

A major advantage of the large-$\beta_0$ result for ${\cal K}$ is that its analytic structure in the Borel plane is known. We will use this structure as a basis for modeling the Borel function in the full theory. We shall require consistency with the first few orders that are known, such that NNLL accuracy is guaranteed. By examining different models we would like to check the sensitivity to yet unknown higher-order coefficients, in particular in the context of open theoretical problems concerning renormalons. There is no attempt here to cover the infinite set of functions that obey the constraints. 

We will assume that the singularity structure of the physical kernel is similar to that obtained in the large-$\beta_0$ limit from~(\ref{large_beta_0_result}) and~(\ref{phys_kernel}). In particular, in our models there will be two renormalons at $u=1$ and $u=2$, but no higher ones. Thus, non-perturbative contributions will be parametrized by (\ref{J_NP}). This is not to say that higher renormalon singularities do not exist in the full theory, but rather that with the information available there is no way we can control their residues. Thus, in our models, similarly to~(\ref{large_beta_0_result}), $B[{\cal K}](u)$ does not vanish at $u=1,2$, but it does vanish at all higher integers. 

We will also restrict ourselves to the case where $B[{\cal K}](u)$ is regular at $u=1,2$ so the renormalon singularities of the physical kernel in our models appear as simple poles. As mentioned in section~2.2, the nature of the singularities is related with the anomalous dimension of higher-twist operators.
Here, we keep $B[{\cal K}](u)$ renormalon free, thus assuming that the anomalous dimension at any twist coincides at large $N$ with that of twist two.
This issue certainly deserves further investigation in the future. 
 
We investigate two types of models for $B[{\cal K}](u)$. The first 
is based on modifying the large-$\beta_0$ function~(\ref{large_beta_0_result})
by a multiplicative factor and the second -- by an additive term.
A~simple example with a multiplicative factor is 
\ba
\label{model_a}
&&\left.B[{\cal K}](u)\right\vert_{{\rm model} \, a}=\frac12 \,\frac{\sin \pi u}{\pi u}\,{\rm e}^{\frac53 u}  \left[\frac1 {1-u}+\frac1{1-u/2}\right] \,\times\,{\rm e}^{u\left(a_2-\frac53\right)}\,(1\,+\,H^{(a)} u^2),\nonumber\\ 
&&H^{(a)} = -\frac34a_2-\frac58-\frac12a_2^2+\frac16\pi^2+\frac12a_3^{\MSbar}
+\frac12\delta_2^{\MSbar}-\frac12\delta a_2-b_2
\ea
where $H^{(a)}$ is fixed to reproduce the exact $\tilde{k}_1$ and $\tilde{k}_2$ of (\ref{kab_tilde_rel}).
Another example (model $b$) is provided by replacing  $(1\,+\,H^{(a)} u^2)$ by  ${\rm e}^{H^{(b)} u^2}$, with $H^{(b)}=H^{(a)}$.
Next, a couple of simple examples with additive (regular) terms are 
\ba
\label{model_c}
&&\left.B[{\cal K}](u)\right\vert_{{\rm model} \, c}=\frac{\sin \pi u}{\pi u}\,\left\{
\frac12 \,{\rm e}^{\frac53 u}  \left[\frac1 {1-u}+\frac1{1-u/2}\right]\,
+\, u\left(a_2-\frac53\right)+u^2H^{(c)}\right\},
\nonumber\\ 
&&H^{(c)} = -\frac{235}{72}+\frac16\pi^2+\frac12 a_3^{\MSbar}+\frac12\delta_2^{\MSbar}-\frac12\delta a_2-b_2
\ea
and
\ba
\label{model_d}
&&\left.B[{\cal K}](u)\right\vert_{{\rm model} \, d}=\,\frac{\sin \pi u}{\pi u}\,\left\{ \frac12 \,{\rm e}^{\frac53u} \, \left[\frac1 {1-u}+\frac1{1-u/2}\right]\,+\,{\rm e}^{u(a_2-\frac53)+u^2 H^{(d)}}-1\right\},\nonumber\\ 
&&H^{(d)} = -\frac{335}{72}+\frac16\pi^2+\frac12 a_3^{\MSbar}+\frac12\delta_2^{\MSbar}-\frac12\delta a_2-b_2-\frac12 a_2^2+\frac53a_2.
\ea
The main difference between the two sets of models, is that models $c$ and $d$ keep the residues of the renormalon singularities at their large-$\beta_0$ values, whereas in models $a$ and $b$ these residues are modified at NLL by purely non-Abelian contributions, proportional to $C_A/\beta_0$, and, starting from NNLL by correlated multi-gluon emission even in the Abelian case. 
In general the residues should be modified by such contributions, but the way the functional form varies is not yet under theoretical control.

\begin{figure}[htb]
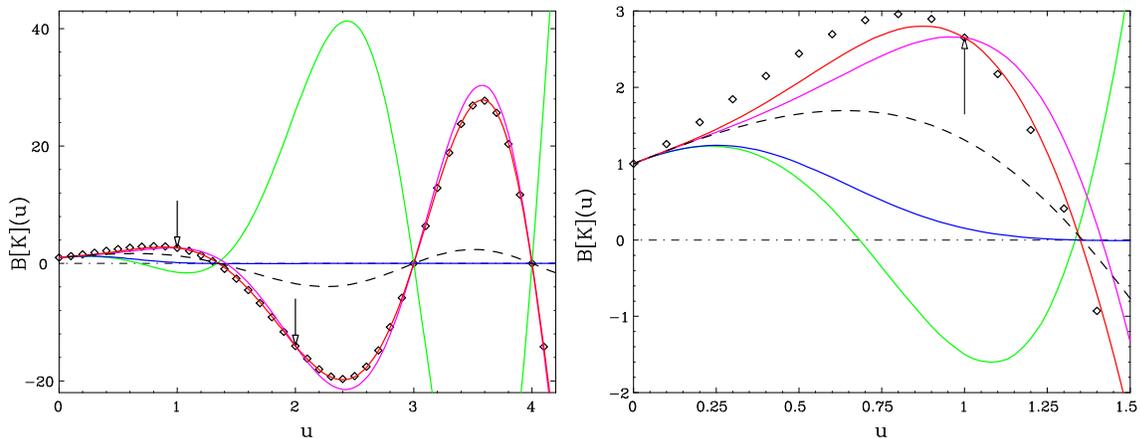

  \begin{center}  
\epsfig{height=7.5cm,angle=90,file=Borel_moedls_large.ps}
\epsfig{height=7.5cm,angle=90,file=Borel_models_small.ps}
\vspace*{-10pt}
  \end{center}
\caption{Different models for $B[{\cal K}](u)$. The points describe the calculated function in the large-$\beta_0$ limit~(\ref{large_beta_0_result}), and the lines describe simple models for this function in QCD, inspired by the large-$\beta_0$ analytic structure, which are consistent with the first three orders ($\tilde{k}_{0,1,2}$) in the expansion
of  $B[{\cal K}](u)$. The full lines stand for the following models (from bottom to top in the range $0<u<1$): green -- $a$, blue -- $b$, magenta -- $c$ and red -- $d$. The dashed line describes model $e$. Arrows show the points where renormalon poles appear.}
\label{models}
\end{figure}

As shown in figure~\ref{models} different models extrapolating from the large-$\beta_0$ limit have very different behaviour indeed. The knowledge of the value of $B[{\cal K}](u)$ at $u=0$ and the first two derivative there cannot constrain it away from this point.    
Note that models with an additive correction ($c$ and $d$) follow closely the 
large-$\beta_0$ functional form away from $u=0$, as their residues at $u=1$ and at $u=2$ are fixed by this limit. On the other hand, models with a multiplicative correction ($a$ and $b$) can have a completely different form and even the sign of the first renormalon residue varies.

\begin{figure}[htb]
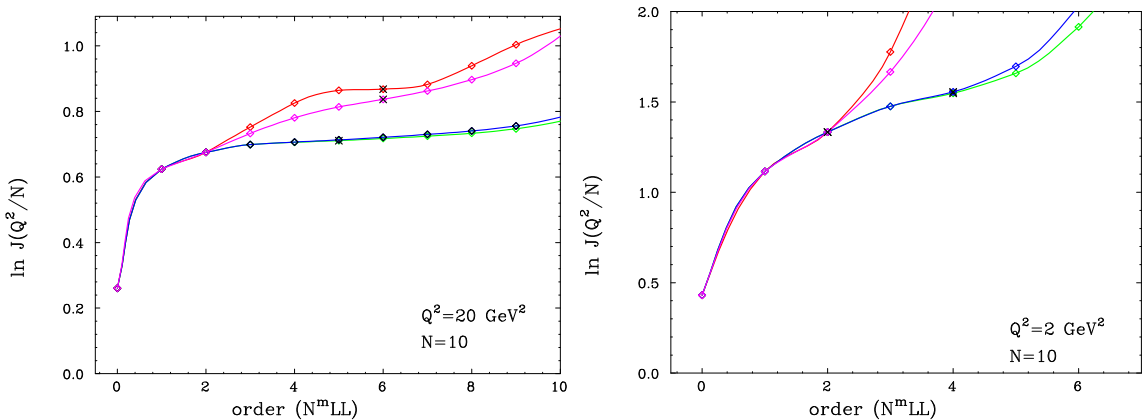

  \begin{center}  
\epsfig{height=7.5cm,angle=90,file=DIS_fla10.ps}
\epsfig{height=7.5cm,angle=90,file=DIS_fla10_10GeV2.ps}
\vspace*{-10pt}
  \end{center}
\caption{Order by order calculation of the jet function $\ln J_N\left({Q^2};\mu_{F}^2\right)$ for $N=10$ and for $Q^2=2$ GeV$^2$ (right box) and $Q^2=20$ GeV$^2$ (left box). NNLL ($m=0,1,2$) are exact in all cases. Subleading logs are calculated based on each of the four models (from bottom to top): green -- $a$, blue -- $b$, magenta -- $c$ and red -- $d$. The minimal term in the series is denoted by an `x' symbol.}
\label{order}
\end{figure}

In general, the detailed behaviour of $B[{\cal K}](u)$ away from the origin, has a minor significance concerning the value of the physical kernel itself, or $J_N\left({Q^2};\mu_{F}^2\right)$. To see the sensitivity to the behaviour of $B[{\cal K}](u)$, we computed $\ln J_N\left({Q^2};\mu_{F}^2\right)$ with the different models described above. The results are shown in Fig.~\ref{order} order by order at increasing logarithmic accuracy.
Differences are rather significant.
Models $a$ and $b$ tend to have a moderate (although not negligible) contributions beyond the NNLL, quite a favourable scenario for perturbation theory. On the other hand, models $c$ and $d$ indicate  huge corrections beyond the NNLL. The resolution of the theoretical questions allowing to pick the right model for subleading logarithmic corrections thus becomes crucial for any analysis oriented at power accuracy. 

One typical feature of models $c$ and $d$ is that contributions of consequential orders are not monotonically decreasing up to the minimal term, but instead, their magnitude oscillates. This, however, should not be too surprising given that in the case under consideration oscillations in magnitude of consequential terms occur at the first few order: the NLL contribution is  larger than the LL one, while the NNLL contribution is smaller than both. 
 
As mentioned above, already when constructing a model for $B[{\cal K}](u)$ which is consistent with NLL accuracy in the non-Abelian theory, there are various possibilities. Nevertheless, in this case it seems natural to use a multiplicative modification of the large-$\beta_0$ result, such that $C=5/3$ at the exponent in~(\ref{large_beta_0_result}) is effectively replaced for the full $a_2$ coefficient~(\ref{MSbar_coef}). 
In fact, the choice made when promoting the large-$\beta_0$ result to NLL accuracy is the most significant one. This can be verified by examining the model,
\be
\left.B[{\cal K}](u)\right\vert_{{\rm model} \, e}=\,\frac{\sin \pi u}{\pi u}\left\{\frac12 \,{\rm e}^{a_2u} \, \left[\frac1 {1-u}+\frac1{1-u/2}\right]\,+\,{\rm e}^{u^2 H^{(e)}}-1\right\}
\label{model_e}
\ee
with $H^{(e)}=H^{(a)}$, where $5/3$ is replaced by the full $a_2$ in the first term, while further corrections (NNLL) are incorporated as an additive term.
This intermediate step between the two classes of models described above, turns out to be quite close to the ones with a multiplicative correction, $a$ and $b$. For example, for $N=10$, $Q^2=20$ GeV$^2$, upon truncation at the minimal term model $a$ yields $\ln J=0.713$, model $d$ yields $\ln J=0.868$ and model $e$ yields $\ln J=0.755$.  

Indeed, in previous applications of DGE~\cite{Thrust_distribution,DGE}, the procedure chosen to extrapolate from the large-$\beta_0$ limit and comply with NLL accuracy was the replacement $5/3\longrightarrow a_2$. This follows the spirit of~\cite{CMW} who first noted the special r\^ole of the ``gluon bremsstrahlung'' effective charge, (defined by the cusp anomalous dimension~${\cal A}$) in this context. This natural generalization is sufficient at NLL, but not beyond it.
In any case, the deeper theoretical problem of how purely non-Abelian contributions as well as correlations in multiple gluon emission modify the renormalon structure of the Sudakov exponent remains open. 

With this caution in mind, we will proceed here on good faith with models with a multiplicative modification of~(\ref{large_beta_0_result}), and in particular, with model $a$, to perform data analysis. A multiplicative modification can always be recast as a modification of the coupling which is integrated over by redefining $T(u)$ in (\ref{J_N}), and thus it can be regarded as the most natural generalization of the large-$\beta_0$ limit result. Based on Fig.~\ref{order} it is also an optimistic choice, leading to rather moderate higher-order corrections. 

\section{Data analysis}

In the previous section we summarized the state-of-the-art knowledge (and ignorance) on the twist-two coefficient function of $F_2$ at large $x$. We also have a picture~\cite{DIS} concerning the structure of higher twist, with a concrete formula for their parametrization (\ref{J_NP}).
We now turn to investigate the implications this has on data analysis.
The following questions will be emphasized:
\begin{itemize}
\item{} Are the data and the theoretical description of $F_2$ consistent?
\item{} Is there any experimental indication for higher-twist contribution to $F_2$?
\item{} To what extent can one determine the relevant parameters from the data. This includes $\alpha_s$, the non-singlet quark distribution $q_N(\mu_F^2)$ and the higher twist parameters $\omega_i$.
\item{} What is the significance of Sudakov resummation as compared with the NNLO result for the coefficient function? 
\item{} What is the significance of renormalon resummation in the Sudakov exponent and the current uncertainty about subleading logs (beyond NNLL)?  
\end{itemize}

In order to systematically treat target-mass corrections and simplify the resummation procedure, we work here in moment space. For this purpose we compute the Nachtmann moments from the available experimental data on $F_2^p(x,Q^2)$.  We also include the
contributions at $x=1$ from the proton electric and magnetic
form factors.
The data we use are from a series of SLAC experiments on $e p$ scattering~\cite{SLAC_ref1}--\cite{SLAC_ref9}
and the BCDMS collaboration using $\mu p$  scattering~\cite{BCDMS}. 
 
Before coming to the fits, let us briefly describe the data and 
the way we use it. 
The SLAC data cover a region from $Q^2$ below 1 GeV$^2$ to
20 GeV$^2$ and $x$ up to $x_{\max}$ defined by the minimum
value of $W^2 = (m_p+m_\pi)^2$. The BCDMS data cover a region
$x \leq 0.75$ for $20<Q^2<230$ GeV$^2$.

The moments are computed in $15$ bins of $Q^2$ from 2 to 105 GeV$^2$.
There is a region of overlap between the SLAC and BCDMS data sets.
While small it does allow a check on the relative normalisation
of the two experiments.
The computation involves estimating the contribution from the
region between the highest $x$ bin at each $Q^2$ value and
$x=x_{\max}$. We take a linear extrapolation with an arbitrary
uncertainty of $100$\%.
For the SLAC data this is a very small correction.
The final errors on the values of each moment include this 
uncertainty.

The physics analysis we apply assumes simply the non-singlet
evolution of moments. This is equivalent to assuming that high enough  moments of $F_2^p(x,Q^2)$ are dominated by just the $u$ and $d$ valence
quarks. This can be checked by taking a standard global fit to the
DIS data~\cite{Martin:2002dr}
 and evaluating the relative contribution to each moment
from the sea and valence quarks. We find at $Q^2$ = 10 GeV$^2$, for
example, that the LO analysis of MRST2001 gives valence/total for
$N=4$ to be over 90\% and for $N \geq 5$ to be more than~95\%. 
We therefore choose to analyse moments $N \geq 5$ and keep in mind the valence approximation as a source of an overall $\lsim 5$\% uncertainty.

\subsection{NNLO based fits}

It is well known~\cite{Martin:1998np} that a NLO analysis of DIS data 
fails to describe
the large-$x$ region. Usually the deviations are attributed to
higher-twist contributions at low $W^2$. In the MRST2001
NLO leading-twist analysis, data from $W^2 \geq 12.5$ GeV$^2$ were 
simply dropped. Even including the NNLO corrections which are now available~\cite{FRS}--\cite{vanNeerven:2001pe},
though helping a little, did not succeed in diminishing the discrepancy at
 large~$x$.
\begin{figure}[htb]
  \begin{center}
\vspace*{-25pt}  
\mbox{\kern-0.5cm \epsfig{height=12.5cm,width=12.5cm,angle=0,file=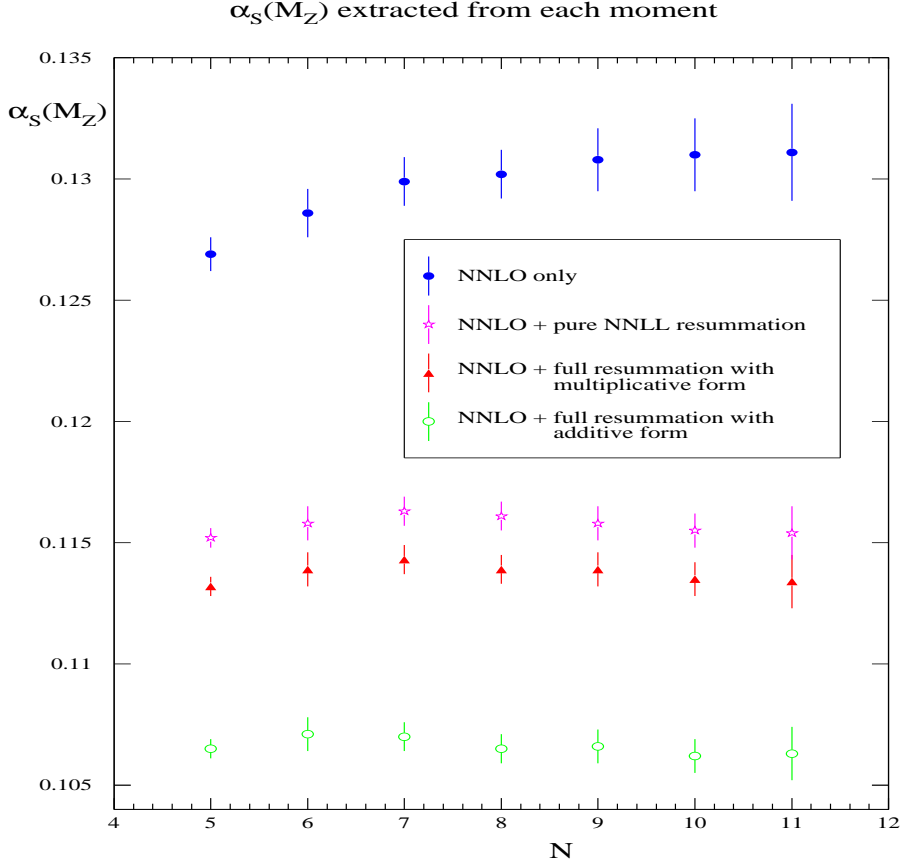}}
  \end{center}
\vspace*{-20pt}
 \caption{The extracted value of the coupling from individual moments for $F_2^p$ in various approximations. The symbols represent (from top to bottom):
a NNLO fit with no Sudakov resummation and three fits with Sudakov resummation, 
a pure NNLL and two renormalon-based models: $a$ -- multiplicative modification of the large-$\beta_0$ limit, and $d$ -- additive modification. }
\label{alpha_s}
\end{figure}

\begin{figure}[htb]
  \begin{center}
\vspace*{-30pt}    
\mbox{\kern-1.5cm
\epsfig{height=15truecm,width=18truecm,angle=0,file=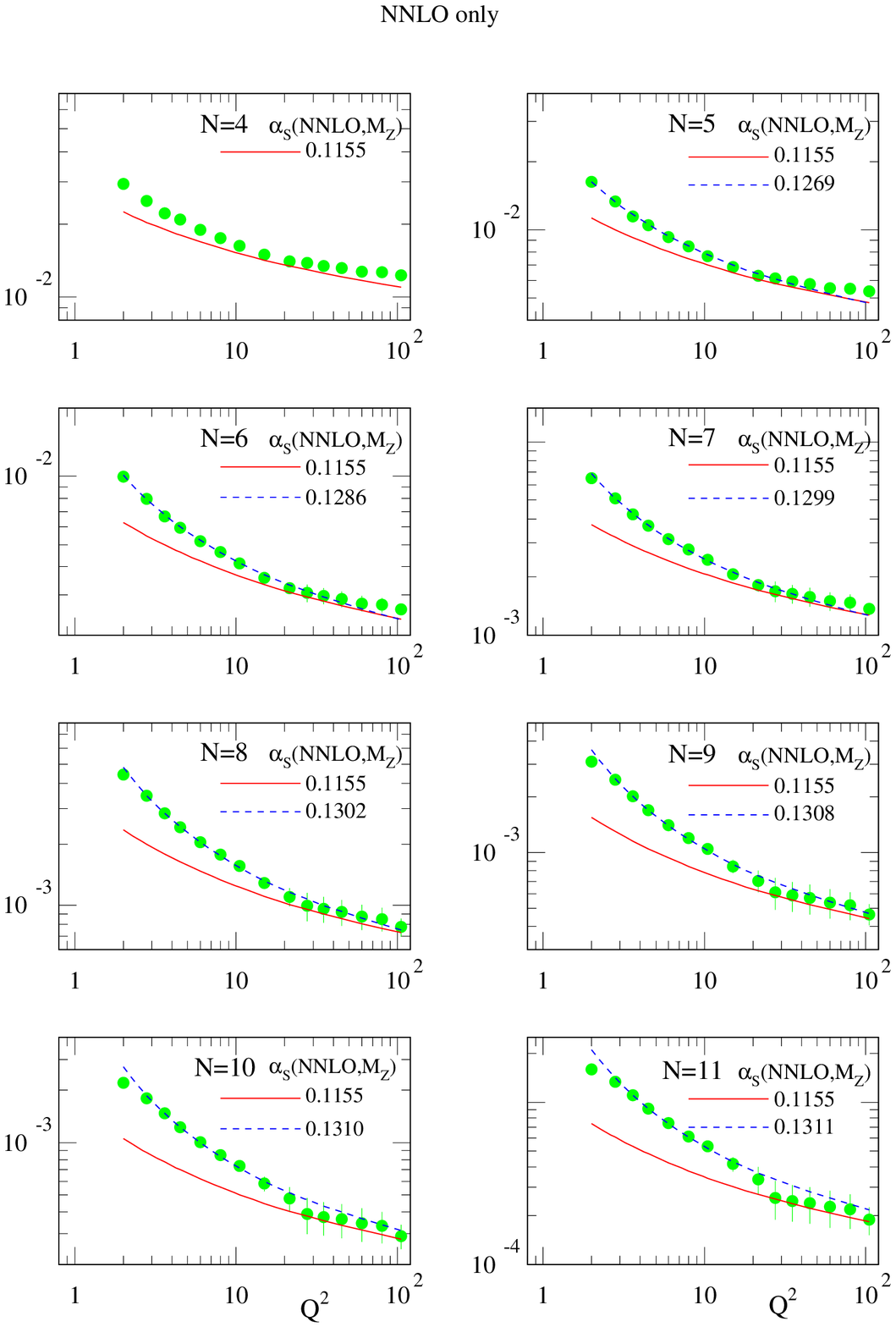}}
  \end{center}
\vspace*{-25pt}
\caption{Experimental moments $N=4$ through 11 of $F_2^p$ with two NNLO descriptions for each $N$: one (dashed line) based on fitting $q_N(\mu_F^2)$ and
$\alpha_s^{\MSbar}(M_Z^2)$ and the second (full line) based on extrapolating the standard MRST fit to low~$W^2$.}
\label{NNLO_comp}
\end{figure}

This situation is confirmed here by attempting to fit the moments we extracted  by a NNLO formula,
\begin{equation}
\label{F2_NNLO}
F_2^N(Q^2) = q_N(\mu_F^2)\, E_N(Q^2,\mu_F^2)\,C_N(Q^2). 
\end{equation}
We use here the standard ${\overline {\rm MS}}$ factorization and renormalization schemes. The NNLO evolution factor is
\be
E_N(Q^2,\mu_F^2)\,=\,\exp\left\{-\frac12\int_{\mu_F^2}^{Q^2}\gamma(\mu^2)d\mu^2\right\}
\,=\,\left [ \frac{\alpha_s(Q^2)} 
{\alpha_s(\mu_F^2)} \right ] ^{d_N}
\,[1 + r_N \alpha_s(Q^2) + s_N \alpha_s^2(Q^2)],
\label{E_NNLO}
\ee
and the NNLO coefficient function with the factorization scale (and renormalization scale) set to $Q^2$ is
\be 
C_N^{\NNLO}(Q^2)=1 + c_N^{(1)}\alpha_s(Q^2) + c_N^{(2)}\alpha_s^2(Q^2).
\label{C_NNLO}
\ee
In (\ref{E_NNLO}),  $d_N=\gamma_N^{(0)}/2\beta_0$, $r_N$ involves the NLO anomalous
dimension $\gamma_N^{(1)}$ and $s_N$ the NNLO anomalous dimension 
$\gamma_N^{(2)}$. We use the results of Retey and Vermaseren~\cite{Larin:1993vu} for the $\gamma_N^{(2)}$ for even $N$ and a smooth interpolation of these for odd~$N$. The running of the coupling $\alpha_s(Q^2)$
is approximated by the NNLO $\beta$ function. 

We fit each moment in turn with two parameters, the moment of
the valence quark distribution $q_N(\mu_F^2)$ and
$\alpha_s^{\MSbar}(M_Z^2)$. 
The resulting values of the latter are shown in Fig.~\ref{alpha_s} 
(upper box) showing two features: (i) large value
of the coupling and (ii) a marked increase as $N$ increases.
The MRST2001 NNLO fit with $W^2 \geq 12.5 $ GeV$^2$ prefers
a much lower value, $\alpha_s^{\MSbar}(M_Z^2)=0.1155$. 
In Fig.~\ref{NNLO_comp} we show two NNLO descriptions, one with the
pairs of parameters just described and the other corresponding to the extrapolation of the MRST2001 NNLO description to low $W^2$.
We conclude, with no surprise, that the leading-twist NNLO formula
cannot satisfactorily describe the large $N$ moments of $F_2^p$.

\subsection{Resummation based fits}

\begin{figure}[htb]
  \begin{center}
\vspace*{-25pt}    
\mbox{\kern-0.5cm
\epsfig{height=13truecm,width=15truecm,angle=0,file=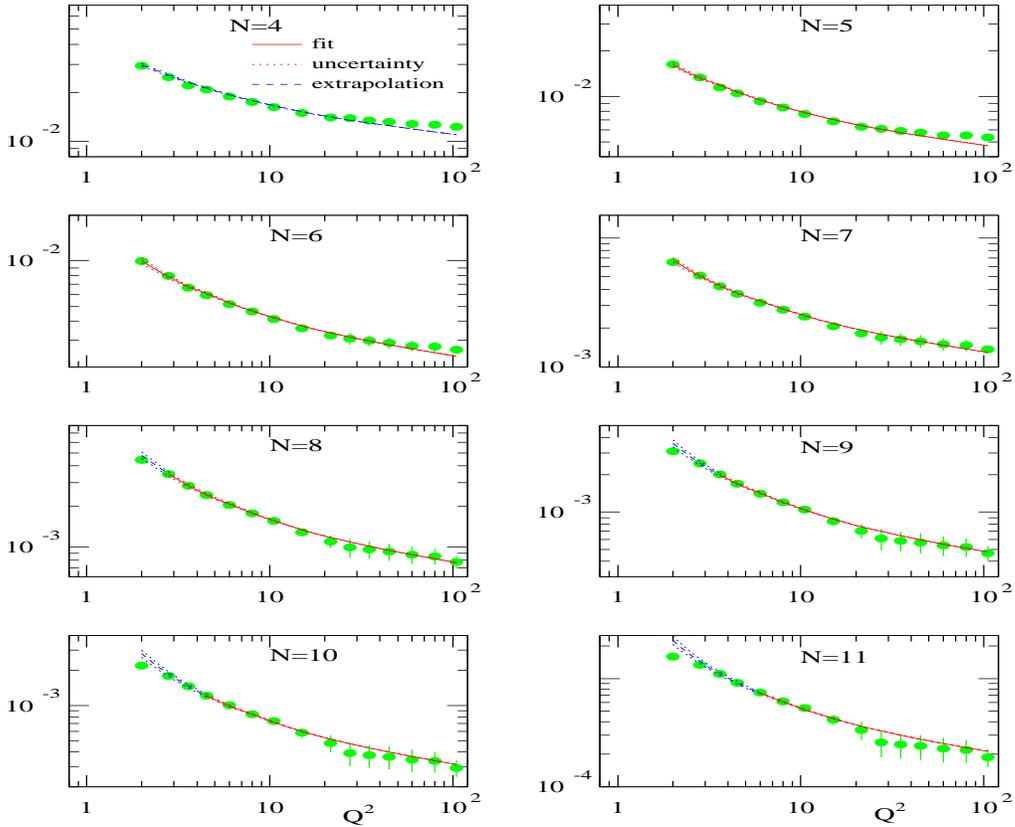}}
  \end{center}
\vspace*{-20pt}
\caption{Experimental moments $N=4$ through 11 of $F_2^p$ and a perturbative description based on Sudakov and renormalon (model $a$) resummed coefficient function,   matched into the NNLO result. Here $\alpha_s^{\MSbar}(M_Z^2)=0.1135$ (this is the best fit value). The full line represents the region on which the fit is based whereas the dashed line represents an extrapolation to lower $W^2$. Dotted lines represent the resummation ambiguity based on ($\pm$) the size of the minimal term.}
\label{mom2}
\end{figure}
\begin{figure}[htb]
  \begin{center} 
\vspace*{-25pt}   
\mbox{\kern-0.5cm
\epsfig{height=13truecm,width=15truecm,angle=0,file=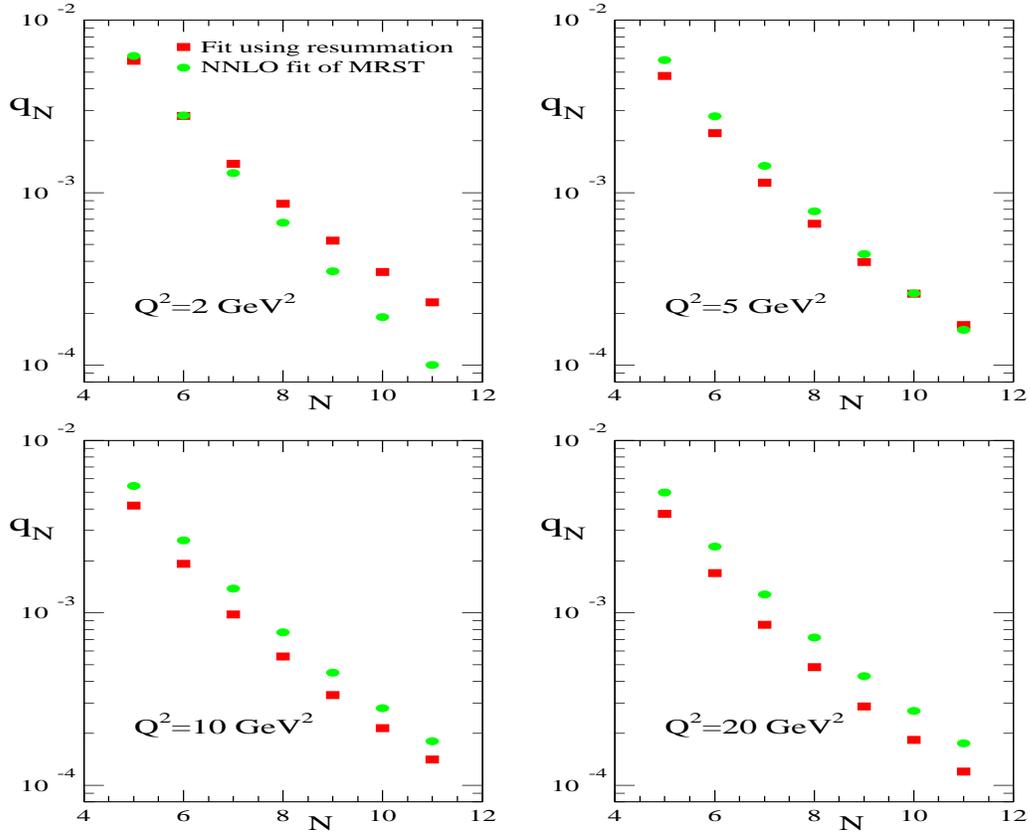}}
  \end{center}
\vspace*{-20pt}
\caption{Values of the parameters $q_N(\mu_F^2)$ for $\mu_F^2=$ 2,5,10
and 20 GeV$^2$. The differences between the values for the two cases
(i) $W^2 \leq 12.5$ GeV$^2$ excluded with NNLO evolution only and (ii)
low $W^2$ data included with NNLO + Sudakov resummation indicate the
modifications required to the large $N$ moments due to resummation effects.}
\label{qnvals}
\end{figure}

Next consider the additional effect of Sudakov resummation. 
This is done by modifying eq.~(\ref{C_NNLO}) to include higher-order terms which are enhanced by powers of $\ln N$. We use, the so-called log-R matching, namely
\be
\ln C_N(Q^2)=\ln  C_N^{\NNLO}(Q^2) +\ln J_N(Q^2)-\ln C_N^{\overlap}
\label{matching}
\ee 
where $\ln J_N(Q^2)$ is given by~(\ref{J_N}) or~(\ref{lnJ_fla}) with the factorization scale set to~$Q^2$. The subtracted term accounts for log-enhanced 
contributions up to order $\alpha_s^2$ which are included in both the NNLO coefficient function and in the Sudakov resummed one,
\ba
\ln C_N^{\overlap}&=&\frac{C_F}{\beta_0}{\bigg [}\left(\frac12 L^2+\left(\gamma_E-b_1\right) L\right) A  
\\\nonumber&&+
\left(\frac16 L^3+\left(\frac12 \gamma_E+\frac12 (a_2-b_1)\right) L^2
+\left(\gamma_E(a_2-b_1)-b_2+ \frac{\gamma_E^2}{2}+\frac{\pi^2}{12}\right) L\right) A^2{\bigg ]}.
\ea

We apply the two following criteria in selecting data for the fits. Firstly,
we fit moments $N \geq 5$ , to ensure non-singlet
dominance. Secondly, we limit
$\lambda = A\ln N= (\beta_0 \alpha_s/\pi) \ln N \leq \lambda_{\max}$, to ensure the validity of the perturbative treatment. 
We choose $\lambda_{\max} = 0.4$. Since $\lambda$
depends on one of our fit parameters, $\alpha_s(M_Z^2)$, the number of
data points included in a fit could vary with this parameter.
To avoid this we choose the data to be fitted for $\lambda_{\max}$
evaluated for the MRST2001 NNLO default value, $\alpha_s(M_Z^2)
= 0.1155$. The moments of the experimental data going into the
fits are given in Table~\ref{edp}. 
\begin{table}
\begin{center}
\begin{tabular}{|l|l|c|}
\hline
$N$&$Q^2_{\min}$&no. of points\\
\hline
5   &   2     &           15         \\ 
6,7,8 &  2.8    &           14         \\
9,10&   3.6   &           13         \\ 
11  &  4.5    &           12         \\
\hline                                        
\end{tabular}
\caption{The experimental data points included in the fits. For all $N$ values $Q^2_{\max}=105$ GeV$^2$. The number of data points with different $Q^2$ values is indicated in the last column.\label{edp}}  
\end{center}
\end{table}
Note that we chose this same data set for the pure NNLO fits for the
sake of comparison. 

As in the NNLO fit we have two free parameters, $q_N(\mu_F^2)$ and
$\alpha_s^{\MSbar}(M_Z^2)$. We choose three variants for Sudakov resummation.
The first is the standard, NNLL accuracy resummation, and the other two include renormalon resummation according to models $a$ and $d$ discussed in section 2.5. Contrary to fixed logarithmic accuracy, renormalon resummation is oriented at power accuracy. The renormalons are regularized here simply by truncating the series~(\ref{lnJ_fla}) at the minimal term.  

The results are shown in Fig.~\ref{alpha_s}. In contrast to the NNLO case, the values of the coupling are practically independent of $N$.
This is true for any of the renormalon-based models as well as for pure NNLL resummation. Thus, Sudakov resummation proves to be important: it provides the ingredient which is missing in the fixed-order perturbative description discussed above.

However, as shown in Fig.~\ref{alpha_s}, the extracted value of $\alpha_s(M_Z^2)$ varies significantly depending on the chosen model. In pure NNLL resummation, typically $\alpha_s(M_Z^2)=0.1155\pm 0.0010$\footnote{The error quoted, like the bars in Fig.~\ref{alpha_s}, reflects just the propagated experimental error (taking into account the uncertainty in extrapolating the data to large $x$). It does not include any estimate of theoretical uncertainty, for example due to factorization or renormalization scale dependence.} while with renormalon resummation $\alpha_s(M_Z^2)$ varies between $0.1135\pm 0.0010$ (model $a$) and $0.1070\pm 0.0010$ (model $d$). 
Since in any case NNLL accuracy is guaranteed, these differences are due to sub-leading logs, N$^3$LL and beyond.
Note that the result quoted for model $a$ actually represents a large class of models with a multiplicative modification of the large-$\beta_0$ all-order result. The details of this factor have a minor effect of the result. Taking the optimistic approach that these multiplicative models represent well the structure of the Sudakov exponent, the central value of the strong coupling from this analysis is $0.1135$. Clearly, more theoretical work is required to reduce the uncertainty.

Fig.~\ref{mom2} shows the results of fitting $q_N$  based on model $a$ for a fixed value of $\alpha_s(M_Z^2)=0.1135$. The plotted curves are almost identical to those obtained in a similar fit where~$\alpha_s$ is free, since the variations of the best-fit value of $\alpha_s$ as a function of~$N$ are small (Fig.~\ref{alpha_s}).
Note that in summing the series~(\ref{lnJ_fla}), the asymptotic expansion is terminated at the minimal term. A sensible estimate of the corresponding power-suppressed ambiguity is simply the size of the last term included in the sum. Fig.~\ref{mom2} shows the magnitudes of this ambiguity together with extrapolations of the fits to
regions excluded from the fitting procedure.  
Under the assumption taken here that the dominant higher-twist terms are those associated with mixing with the leading twist, this power-suppressed ambiguity should provide an estimate of the magnitude of the higher twist. As clearly seen in Fig.~\ref{mom2} this is a small correction even in the highest moment.  
These results imply that the low $W^2$ data, i.e. $W^2 \leq 12.5$GeV$^2$ 
can be successfully described upon improving the 
standard NNLO description by including Sudakov resummation. The resulting values
of $\alpha_s^{\MSbar}(M_Z^2)$ are stable with $N$ and are close to the
value extracted from the MRST NNLO analysis where a cut on low $W^2$
is  imposed.

The other parameter $q_N(\mu_F^2)$ is shown in Fig.~\ref{qnvals}
for four values
of $\mu_F^2$ for comparison with the corresponding values expected
from the MRST NNLO analysis. For the 
most part the ranges of $x$ being explored in this analysis of the
high moments correspond to ranges of $W^2$ below the value 12.5 
GeV$^2$ excluded by the MRST analysis.  Thus Fig.~\ref{qnvals} is an indication
of how much the large $x$ parton distributions would have to be modified
to take account of this resummation effect.

\subsection{Fitting the higher twist}

\begin{figure}[htb]
  \begin{center}  
\vspace*{-50pt}  
\mbox{\kern-0.5cm
\epsfig{height=13truecm,width=15truecm,angle=0,file=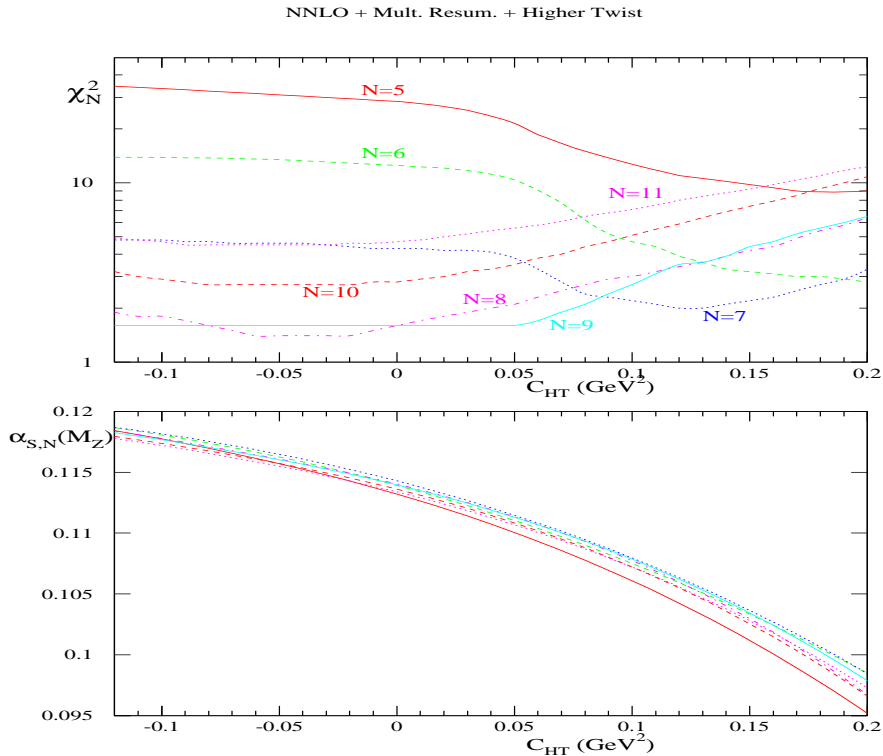}}
  \end{center}
\vspace*{-20pt}
\caption{The total $\chi^2$ as a measure of the quality of the fits (upper box) to individual moments when $C_{\HT}$ is varied and $q_N(\mu_F^2)$ and  $\alpha_s(M_Z^2)$ are free parameters, and the corresponding best-fit value of $\alpha_s(M_Z^2)$ (lower box). Model $a$ is assumed.}
\label{ht}
\end{figure}

Finally, according to eq.~(\ref{J_NP}), we include the simplest model for higher-twist corrections through an extra factor ${J}^{\NP}(N \Lambda^2/Q^2)$ where
\begin{equation}
{J}^{\NP}(N \Lambda^2/Q^2) = \exp {\left\{\frac{C_{\HT} N}{Q^2}\right\}}, 
\end{equation}
with $C_{\HT}\equiv -\omega_1\frac{C_F}{\beta_0}\,\Lambda^2$. Here we neglect 
subleading corrections proportional to $\omega_2$.

We perform fits on the same set of moments as before, but with the 
extra parameter $C_{\HT}$ characterising the additional $Q^2/N$
dependence from the higher twist. Naturally, there is a
compensation between the magnitude of the parameters 
$\alpha_s(M_Z^2)$ and $C_{\HT}$ which is shown in Fig.~\ref{ht} in the lower box. 

For a wide range of $C_{\HT}$ there is little difference in the 
quality of the fits, see the upper box in Fig.~\ref{ht}. The conclusion is clear: there is no phenomenological evidence for a non-vanishing twist-four (or higher twist) contributions going beyond the resummed perturbation theory. 
On the other hand a positive $C_{\HT}$ would decreases the already `lowish' value of $\alpha_s$. 

\section{Conclusions}

In this paper we investigated the large-$x$ limit of the structure function $F_2(x,Q^2)$, concentrating on the interplay between Sudakov resummation, 
renormalons and higher twist. The main assumption we make~\cite{DIS} is that the dominant higher-twist contribution in this region is related to the mixing of higher-twist operators with the leading twist. Consequently, our investigation 
focuses on constraining the twist-two coefficient function at large-$x$.

Our first point is that the physical kernel~$d\ln F_2^N/d\ln Q^2$ receives log-enhanced contributions from a single source, the jet function depending on $W^2=Q^2(1-x)/x$. This property is a consequence of factorization and it  distinguishes structure functions from less inclusive quantities, such are event-shapes, which have an additional observable-dependent sensitivity to large angle soft emission.

Much is known about the structure of the Sudakov exponent of~$F_2$ from general considerations, from the large-$\beta_0$ limit and from fixed-order calculations in dimensional regularization. Nevertheless, the investigation of Sec.~2.5 reveals that subleading logs, N$^3$LL and beyond, which are not yet under theoretical control, may have a large effect. Further theoretical input on the all-order structure of the exponent is required to constrain these contributions. 

While a pure NNLO formula without Sudakov resummation does not describe the large-$x$ data, once Sudakov resummation (NNLL) is employed the {\em purely perturbative} description of $F_2$ is fully consistent with the data. This is reflected in the good fit of the $Q^2$ dependence of each of the moments and in the stability of the extracted $\alpha_s$ as a function of~$N$. This conclusion holds independently of whether renormalons are resummed and of what one assumes about subleading logs.

The theoretical uncertainty concerning the renormalon structure of the Sudakov exponent beyond the large-$\beta_0$ limit, as reflected in the models investigated in Sec.~2.5 and~3.2, translates into an uncertainty of order $\pm 6$\% in the extracted value of the coupling. Thus, taking a conservative approach one would assign a rather large uncertainty to $\alpha_s$ from scaling violation of $F_2^N(Q^2)$ with $N\gsim 5$. A more optimistic approach would be 
to assume that a multiplicative modification of the large-$\beta_0$ result, such as model $a$ in Eq.~(\ref{model_a}), represents well the all-order structure of the exponent. In this case the total effect of subleading logs, N$^3$LL and beyond, is similar to the NNLL contribution (see Fig.~\ref{models}). As a result the central value of $\alpha_s(M_Z^2)$ changes from the pure NNLL result of $0.1155$ to $0.1135$.

A priori, higher twist is expected to be important for $F_2$ at large $x$. However, as it stands, there is no experimental evidence for a non-vanishing higher-twist contribution going beyond the resummed perturbation theory. 
This is consistent with (at least, it does not contradict) the assumption of~\cite{DIS} that the dominant higher twist at large-$x$ is the one related to mixing with the leading twist. 
On the other hand, in the experimentally accessible range and excluding the very small $Q^2/N$ region ($Q^2/N\sim \Lambda^2$), the natural size of the higher twist under the above assumption is still significantly smaller than the theoretical uncertainty.

In addition to solving the theoretical problems mentioned above, progress in this field requires additional data. Although the SLAC and BCDMS data are reasonably precise it is desirable to extend the range they cover. Ideally, to verify or falsify Eq.~(\ref{non_PT_fact}) one would need data covering a large range of $N$ and $Q^2$ with a fixed $Q^2/N$.

\vspace*{30pt}
\noindent{\large {\bf Acknowledgements}} \\ \\

\noindent
We would like to thank Volodya Braun, Stefano
Catani, Andreas Freund, Gregory \hbox{Korchemsky}, Douglas Ross and Bryan Webber for useful discussions. EG thanks the DFG for financial support. RGR thanks the
Leverhulme Trust for the award of an Emeritus Fellowship at~RAL.

\end{document}